\documentclass[12pt]{article}

\usepackage{amsmath}
\usepackage{a4wide}
\usepackage{axodraw}
\usepackage{epsfig}
\usepackage{amssymb}

\newcommand{\pmu}{\partial_\mu}
\newcommand{\pmp}{p_\mu^\pi}
\newcommand{\qmp}{q_\mu^\pi}
\newcommand{\pmk}{p_\mu^K}
\newcommand{\qmk}{q_\mu^K}

\newlength{\abstwidth}
\begin{document}
\begin{titlepage}
  \sloppy
  
  \pagestyle{empty}
 
  \begin{flushright}
    LU TP 04 - 22 \\
    May, 2004\\
    v2, June 2004
  \end{flushright}
  \vfill
  
  \begin{center}
    {\Huge\bf Estimating the Electromagnetic Chiral Lagrangian Coefficients}
\\[15mm]
    {\large  Master of Science Thesis by Anders Pinzke}\\ [2mm]
    {\large  Thesis advisor: Johan Bijnens}\\ [2mm]
    {\it Department of Theoretical Physics,}\\[1mm]
    {\it Lund University, Lund, Sweden}
  \end{center}

  \vspace{\fill}
  \begin{center}
    {\bf Abstract}\\[2ex]
    \begin{minipage}{\abstwidth}
     In the low energy region chiral perturbation theory including virtual
     photons is used to derive the structure of the generating
     functional. The work we do is performed within the three flavor framework and
     reaches up to next-to-leading order. An Euclidian cut-off is
     introduced to separate the long- and short-distance contributions. The
     long-distance part is evaluated in the ChPT framework up to
     $\mathcal{O}(p^4)$. The short-distance part is achieved partly from the
     work of J. Bijnens and J. Prades \cite{htdtemc2}, through perturbative
     QCD and factorization. A matching is made and the finite parts of
     the low-energy constants (LECs) is determined by using already existing data of
     the QCD LECs up to order $p^4$ \cite{htdtemc2},
     \cite{a1}, \cite{a2}.
    \end{minipage}
    \vfill
        {\includegraphics[height=4cm]{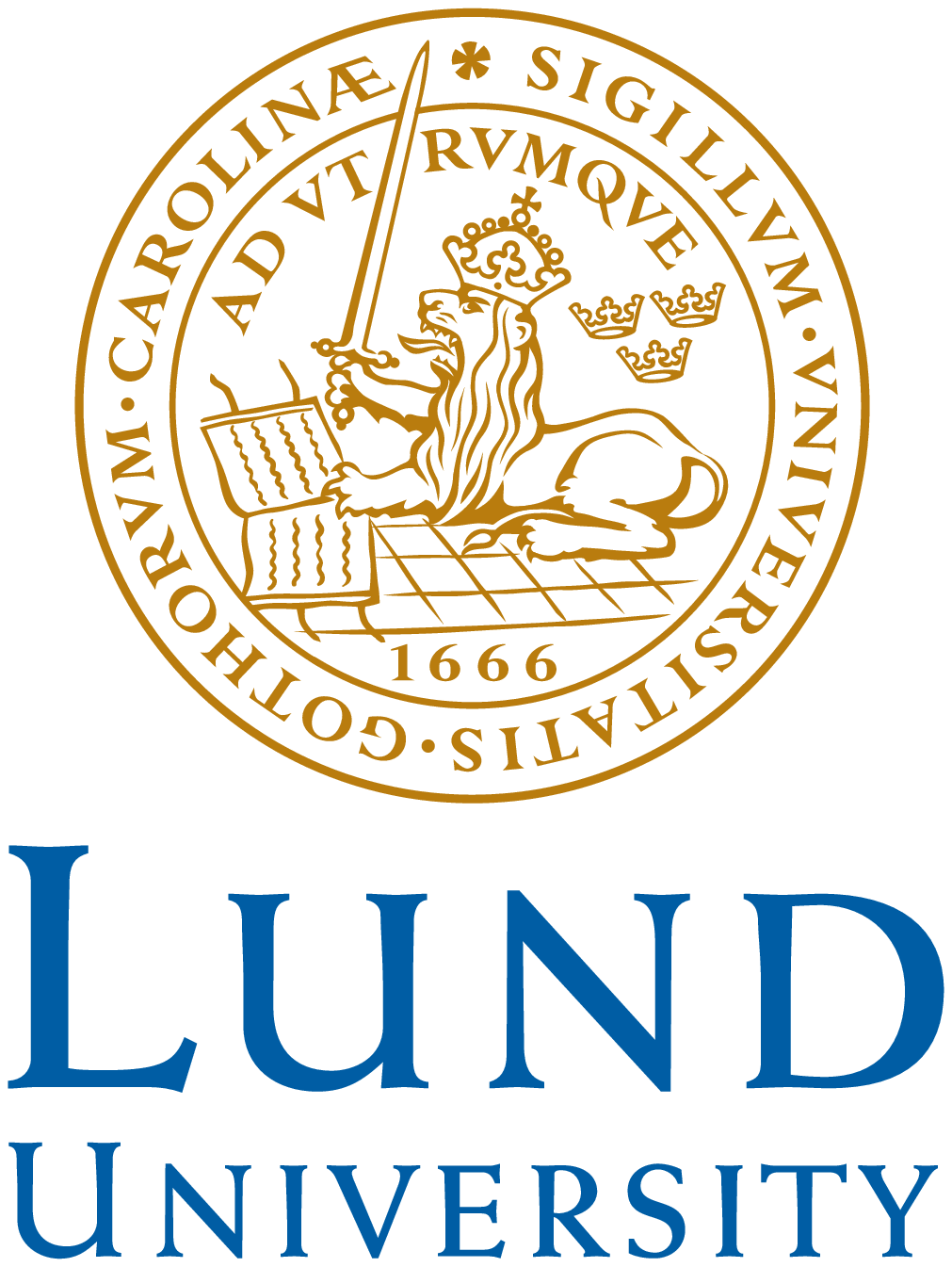}}
\end{center}
  
  \vspace{\fill}
  
  \clearpage
  
\end{titlepage}

\tableofcontents
\clearpage

\section{Introduction}


Masses of hadrons have always been a problem to calculate throughout the
history of modern particle physics. The problem lies in the fact that the
particles are believed to have three different mass contributions; the free
quark mass from the Higgs mechanism, the constituent mass and perturbative
interactions with $\gamma$, W and Z. The last contribution is the one that we are 
concentrating on. These are called radiative corrections and involve interactions
from gauge bosons. For the light mesons that we are going to investigate it is enough
to use photon corrections, in an effective low energy model. This model is
derived from the Standard Model (SM) which is a relativistic quantum field
theory that give rise to
the strong interaction described by Quantum Chromo Dynamics (QCD) and the
electroweak interaction described by Quantum Flavor Dynamics (QFD). QCD works
very well and agrees with experiments at high energies and short distances. It
is much more difficult to examine QCD at low energies and long distances. One
of the most successful approaches is based on the fact that QCD has a 
$SU(n_F)_L\times{SU(n_F)_R}$ chiral symmetry for $n_F$ massless flavors. To
get an effective field theory model
for low energy, a perturbation of the symmetry above is performed. The
perturbation is called chiral perturbation theory (ChPT). 
This theory can be described by certain fields which correspond to the low energy
spectrum of light pseudoscalar mesons, the  $\pi$, $K$, and $\eta$ octet.
When $\mathcal{L}_{EM}$ is added to
$\mathcal{L}_{QCD}$ the effective Lagrangian for QCD and
Electromagnetic (EM) interactions using Weinberg's power counting scheme
\cite{i2} can be written as:
\begin{equation}
  \label{eqi1}
  \mathcal{L}_{effective}=
  \mathcal{L}_{2}+\mathcal{L}_{4}+\cdots=
  \mathcal{L}_{2}(C, F_0, B_0)+\sum_{i=1}^{10}L_iO^i_4+
  \sum_{i=1}^2H_i\widetilde{O}^i_4+\sum_{i=1}^{14}K_i\widehat{O}^i_{4}+\cdots
\end{equation}
where the subscript has been ordered in momenta, mass and electric charge. The $O^i$s 
represent the term in the effective Lagrangian that are connected to the i:th LEC
(see Eq.~\eqref{eqntloel1}).
The lowest order Lagrangian is given by the non-linear sigma model coupled to
external fields \cite{i1}\cite{ntloel1}. It contains three free parameters in the
chiral limit: the scalar quark condensate $B_0$, the pion-decay constant $F_0$ and 
the lowest order EM coupling constant C. The leading order expansion
$\mathcal{O}{(p^2)}$ implies tree level diagrams with 
$\mathcal{L}_{2}$ vertices. The next-to-leading order expansion is used to
derive the electromagnetic LECs $K_i$ by computing photon loops from
the leading order Lagrangian $\mathcal{L}{_2}$ and adding contributions from
the next-to-leading order Lagrangian $\mathcal{L}_4$. Doing the loop calculations will 
involve all the
29 LECs\footnote{There exist a few more constants up to order $\mathcal{O}(p^4)$, 
but these are of no importance to us} in Eq.~\eqref{eqi1} that are connected to the 
couplings of different vertices in the diagrams, more about this in section 2.5. This 
is not a problem to us since the purely QCD part connected to 
$B_0$, $F_0$, $L_i$ and $H_i$, have been determined experimentally or theoretically. 
The constants of 
order $\mathcal{O}(p^4)$ contain a divergent part and a constant finite part. The 
divergent part can be isolated with the help of dimensional regularization and 
$\overline{MS}$ regularization. For the $K_i$ part this was recently done by Urech in 
\cite{ntloel2}, but the finite
part of the $K_i$ is almost entirely unknown. The goal of this thesis is to
determine the $K_i$ as a function of the $L_i$ and $H_i$. 

\section{Theory}

One of the greatest achievements in physics during the 20th century is the Standard 
model of particle physics. It revolutionized the view of elementary particles. Protons
and neutrons that build up atoms (and electrons, of course) are not considered
to be the fundamental particles any more. Instead they consist of
point-like particles called
quarks. Beside the quarks exist fundamental particles called
leptons. In total there are six quarks (u, d, s, c, b, t) and six leptons ($e$,
$\nu_e$, $\mu$, $\nu_\mu$, $\tau$, $\nu_\tau$) which all matter is
composed of. The forces between the fundamental particles are
described by the twelve gauge bosons. These particles mediate three forces;
the strong-, weak- and the electromagnetic force. Note that the fourth fundamental
force (out of the four known) gravity is not included. The weak force which
comes from the local $SU(2)$ gauge invariance of the SM Lagrangian and the
EM force which originates from the local $U(1)$ gauge symmetry is unified to the
electroweak force (described by QED) which is mediated through the four gauge
bosons ($W^\pm, Z^0, \gamma$). This unified force is based upon the
$U(1)\times{SU(2)}$ symmetry. The strong force comes from the local 
$SU(3)$ gauge symmetry of the Lagrangian. Due to the
non-Abelian\footnote{Whenever the order of transformations matters, they are
  called non-Abelian transformations. The non-zero value of a commutator
   including the Pauli matrices that generate the local SU(2) gauge theory is
   an example of this.}
 nature of the local
$SU(3)$ gauge group, the eight gauge bosons (the gluons) that mediate the
strong force, carry "colour" charges such that the QCD Lagrangian incorporates
gluon self interaction involving vertices with three and four gluons. This
fact complicates things a great deal.

\subsection{Interpretation of the Lagrangians}

The theory of the SM is written in terms of Lagrangians, which describe the dynamics
of a system in a fundamental and well known way. They are scalars in every 
relevant space and invariant under transformations. In the full field theory, the
Lagrangians can be used in a variety of ways. 50 years ago Feynman created a 
simple method which reduces and simplifies amplitude calculations done in 
relativistic quantum field theory. This was an important discovery because the 
amplitude is of great importance since it
incorporates the physical picture of a system. Feynman's method is based on
diagrams (called Feynman diagrams) which can be used to calculate the
amplitude in an efficient way, using so-called Feynman rules. The rules enforce
conservation of energy and three-momentum at each vertex, so that the four-momentum
for the whole vertex is conserved.

Let us investigate an easy example where, $\pi^0\rightarrow \gamma \gamma$. 
The pion, using standard quantum field notation, is written as
\begin{align}
  \pi^0(x)=\int{N_{\pi^0}(p)(a_pe^{-ip\cdot x}+
    {a_p^\dagger}e^{ip\cdot x})}d^3p
\end{align}
where the creation $a^\dagger$ and annihilation $a$ operators, create and
annhilate a pion with momentum p, respectively. The $N_{\pi^0}(p)$ is a
numerical factor determined from the Feynman rules.  
The photon field can also be described in a simular way, but since it is a
vector and not a scalar it needs an extra index. A polarization vector
$\epsilon$ with four components is introduced. The photon field is then
represented by
\begin{align}
  A_\mu (x)=\int{N_A(q)(\epsilon _\mu a_ke^{-ik \cdot x}+
    {\epsilon ^*_\mu a_k^\dagger}e^{ik\cdot x})}d^3q
\end{align}
where k is proportional to the momentum q of the created photons.
For the $\pi^0 \rightarrow \gamma \gamma$ where one pion is annihilated and two
photons are created, the corresponding terms are picked out and used in the
Lagrangian to get the Feynman amplitude for the vertex.

\subsection{Renormalization}

Start by considering lowest order ("tree level") Feynman
diagrams, where $A+A \rightarrow B+B$.
Assume that the time axis goes from the bottom to the top.
\begin{center}
  \begin{picture}
    (150,40)(0,0)
    \SetWidth{1.}

    \Line(5,0)(40,20)
    \Line(5,40)(40,20)
    \put(40,20){\circle*{4}}
    \Text(23,40)[]{B}
    \Text(23,0)[]{A}

    \DashLine(40,20)(120,20){3}
    \Text(80,27)[]{C}

    \Line(120,20)(155,40)
    \Line(120,20)(155,0)
    \put(120,20){\circle*{4}}
    \Text(137,40)[]{B}
    \Text(137,0)[]{A}
  \end{picture}
\end{center}
The diagram has two vertices, so the matrix element $\mathcal{M}$ is
proportional to the coupling constant squared ($g^2$). This diagram is not
the only one that describes this process, e.g the four-vertex diagrams below
also contribute to the process.
\begin{center}
  \begin{picture}
    (350,40)(0,0)
    \SetWidth{1.}

    \Line(5,0)(40,20)
    \Line(5,40)(40,20)
    \put(40,20){\circle*{4}}
    \Text(23,40)[]{B}
    
    \DashCArc(20,10)(7,30,210){3}
    \Text(8,-5)[]{A}
    \Text(23,0)[]{B}
    \Text(35,10)[]{A}
    \Text(8,20)[]{C}
    \put(27,12){\circle*{4}}
    \put(15,6){\circle*{4}}

    \DashLine(40,20)(120,20){3}
    \Text(80,26)[]{C}

    \Line(120,20)(155,40)
    \Line(120,20)(155,0)
    \put(120,20){\circle*{4}}
    \Text(137,40)[]{B}
    \Text(137,0)[]{A}   
    \Line(175,0)(210,20)
    \Line(175,40)(210,20)
    \put(210,20){\circle*{4}}
    \Text(193,40)[]{B}
    \Text(193,0)[]{A}

    \Text(197,18)[]{C}
    \Text(303,18)[]{C}
    \put(192,10){\circle*{4}}
    \put(308,10){\circle*{4}}

    \DashLine(192,10)(308,10){3}
    \DashLine(210,20)(290,20){3}
    \Text(250,27)[]{A}
    \Text(250,0)[]{B}

    \Line(290,20)(325,40)
    \Line(290,20)(325,0)
    \put(290,20){\circle*{4}}
    \Text(310,40)[]{B}
    \Text(310,0)[]{A}
  \end{picture}
\end{center}
\begin{center}
  \begin{picture}
    (350,60)(0,0)
    \SetWidth{1.}

    \Line(5,0)(40,20)
    \Line(5,40)(40,20)
    \put(40,20){\circle*{4}}

    \Text(12,42)[]{B}
    \Text(33,32)[]{A}
    \DashLine(20,10)(20,30){3}
    \Text(12,-2)[]{A}
    \Text(33,8)[]{B}
    \Text(12,20)[]{C}
    \put(20,10){\circle*{4}}
    \put(20,30){\circle*{4}}

    \DashLine(40,20)(120,20){3}
    \Text(80,26)[]{C}

    \Line(120,20)(155,40)
    \Line(120,20)(155,0)
    \put(120,20){\circle*{4}}
    \Text(137,40)[]{B}
    \Text(137,0)[]{A}   
    \Line(175,0)(210,20)
    \Line(175,40)(210,20)
    \put(210,20){\circle*{4}}
    \Text(193,40)[]{B}
    \Text(193,0)[]{A}

    \DashLine(210,20)(240,20){3}
    \Text(225,26)[]{C}
    \put(260,20){\circle*{4}}
    \GCirc(250,20){10}{1}
    \DashLine(260,20)(290,20){3}
    \Text(275,26)[]{C}
    \put(240,20){\circle*{4}}

    \Line(290,20)(325,40)
    \Line(290,20)(325,0)
    \put(290,20){\circle*{4}}
    \Text(310,40)[]{B}
    \Text(310,0)[]{A}
  \end{picture}
\end{center}
Let us investigate the bottom right diagram containing a loop integral where the
four dimensional element could be written as $d^4q=q^3 dq d\Omega$, where $d\Omega$
represents the angular part. At large $q$ the integrand is essentially
1/$q^4$, so the loop integral has the form of:
\begin{equation}
\int^{\infty}\frac{1}{q^4}q^3dq=\ln q\mid^\infty = \infty
\end{equation}
This integral is logarithmic divergent for large $q$, which was a problem
for a long time. Now there exist systematic methods to get rid of the "problem". 

The procedure goes as follows:
\newline
(1) \emph{Regularize the integral}: Done by modifying the theory, in such a way 
that it remains finite and well-defined. This can be done in a couple of different 
ways which all lead to the same physical result. In this thesis we will use two 
different methods, 
the cut-off procedure and dimensional regularization. The first method mentioned, 
is the oldest one of the two and it is often used on simpler cases. 
The other procedure, dimensional regularization, is a powerful method which is 
based on modifying the dimensionality of the integrals so that they become finite.
\newline
(2) \emph{Renormalize}: Where the physical parameters of interest change from the ones 
introduced by original Feynman rules, to ``renormalized'' and finite parameters 
containing extra factors.
\newline
(3) \emph{Eliminate the regularizing parameters}: One revert from the regularized 
theory back to QED, i.e. the original infinities of QED relate the original (bare) 
particles with the physical particles. This means that the physical observations of 
the theory expressed in terms of mass and charge are finite as one restores QED.

We will now demonstrate the cut-off procedure to give a feeling about the whole 
topic. So, the first step is to regularize the integral, where a suitable
cut-off procedure is used that removes the infinities for the moment. This renders 
the integral to a finite value without spoiling the Lorentz invariance. The cut-off
 is introduced as a ``fudge factor'' that goes to 1 as the cut-off 
$\rightarrow \infty$. The nice thing is that the integral can then be calculated and 
separated into two parts: a finite part,
independent of the cut-off, and a term related to the logarithm of the cut-off,
which diverges when the cut-off goes to infinity. Now something extraordinary
happens, all the divergent cut-off dependent terms appear together with the
masses and coupling constants. This means that the physical masses and
coupling constants are not the m's and g's from the original Feynman rules,
instead they are considered to be renormalized ones, i.e
\begin{align}
&m_{physical}=m+\delta m
&&g_{physical}=g+\delta g
\label{eqr1}
\end{align}
In the infinite cut-off limit the delta factors are infinite, but this
does not matter, since they are not the measured ones. The measured
$m_{physical}$ and $g_{physical}$ are finite, so the m and g in Eq.~\eqref{eqr1} also 
diverge, and makes the right side contribution finite. The remaining finite (cut-off
independent) terms from the loop integrals, also suffer from the modifications
of m and g. But the change in these finite effective masses and coupling constants
also depend on the momentum of the particles involved, and therefore they are called
running masses and running coupling constants. If all the infinities from higher 
order diagrams behave in the way described
above, we say that the theory is renormalizable. 30 years ago 't Hooft showed
that all gauge theories, including QCD and QED, are renormalizable \cite{r1}.

Another interesting way of viewing renormalization is with the help of a medium. 
From quantum 
physics we know that every particle moving through a medium consisting of quantum 
fluctuations which effect values i.e. charge and mass. To be able to ignore the medium 
the values of the particles parameters have to be changed to scale-dependent values, 
and you say that the parameters have been renormalized by the medium.




\subsection{Symmetries}

Symmetries are very important in particle physics. Whenever there is an
invariance there is a corresponding conserved quantity. That is exactly what
Noether's theorem says: Symmetry $\Leftrightarrow$ conserved quantity. These
conserved quantities and the associated currents create, with the help of group
theory the building blocks of the standard model. 
The SM is full of different symmetries. It is built up out of the three
local gauge symmetries $SU(3)_C\times{SU(2)_L}\times{U(1)_Y}$, some continuous
global symmetries and some discrete symmetries.

The full QCD Lagrangian with the $SU(3)_C$ gauge symmetry can be approximated
by a light-flavor Lagrangian, because the scale of interest is $\sim1 GeV$. So
the c, b, and t quark will not be considered, since they are created at a
scale over $\sim1 GeV$. The light-flavor QCD Lagrangian can be
written as 
\begin{align}
  \label{eqs1}
  \mathcal{L}_{QCD}=
  \sum_{q=u,d,s}i\bar{q_L}D \hskip-0.6em/ q_L+
  i\bar{q_R}D\hskip-0.6em/ q_R-m_q(\bar{q_L}q_R+
  \bar{q_R}q_L)
\end{align}
where $q_{L, R}=\frac{1}{2}(1 \pm \gamma_5)q$,
$D\hskip-0.6em/=i\gamma^\mu(\partial_\mu-iG_\mu)$ and $G_\mu$ the gluon
field. The q corresponds to a
particular quark state, and its wave function is a product of factors,
\begin{equation}
  q=
  \left (\begin{array}{c}\mathrm{space}\\\mathrm{factor}\end{array}\right) \times
  \left (\begin{array}{c}\mathrm{spin}\\\mathrm{factor}\end{array}\right) \times
  \left (\begin{array}{c}\mathrm{U(1)}\\\mathrm{factor}\end{array}\right) \times
  \left (\begin{array}{c}\mathrm{SU(2)}\\\mathrm{factor}\end{array}\right) \times
  \left (\begin{array}{c}\mathrm{colour}\\\mathrm{factor}\end{array}\right)
\end{equation}
where each factor represents some labels, coordinates, or indices. If the
masses in the approximated light-flavor Lagrangian are put equal, the Lagrangian
gets an extra global symmetry. The symmetry is an $SU(3)$ flavor symmetry 
(rotation in quark flavor space).
In the chiral limit the quark masses $m_q$=0, which implies that 
$SU(3) \rightarrow {SU(3)_L\times{SU(3)_R}}$ since the left and right handed 
quarks decouple from each other. So
$\mathcal{L}_{QCD}^{m=0}$ is invariant under $SU(3)_L\times{SU(3)_R}\times
{U(1)_A}\times{U(1)_V}\equiv J$. If we define 
\begin{align}
  \Psi= \left( 
    \begin{array}
      {c} q_u \\ q_d \\ q_s
    \end{array} \right)
\end{align}
then the transformations associated with $J$ symmetry correspond to
\begin{align}
\label{eqs2}
\nonumber\\&SU(3)_{V=L+R}          && \Psi
\rightarrow{e^{-i\vec{\theta}\cdot\vec{\lambda}/2}\Psi}
\nonumber\\&U(1)_A                 && \Psi \rightarrow{e^{-i\epsilon\gamma_5}\Psi}
\nonumber\\&U(1)_V                 && \Psi \rightarrow{e^{-i\epsilon}}\Psi
\nonumber\\&SU(3)_L                && \Psi_L \rightarrow g_L{\Psi_L}
\nonumber\\&SU(3)_R                && \Psi_R \rightarrow g_R{\Psi_R}
\nonumber\\&SU(3)_L\times SU(3)_R  && \Psi_L+\Psi_R \rightarrow g_L \Psi_L +
g_R \Psi_R &&& g_{L,R}{\in}SU(3)_{L,R}
\end{align}
where $\vec{\lambda}$ is a vector with the Gell-Mann matrices as its
elements. These transformations are later on used to derive the structure of
the effective Lagrangians.

A symmetry is usually implemented in one of the two following ways: The Weyl
mode and the Goldstone mode. In the Weyl mode the Lagrangian and the vacuum
are both invariant under a set of symmetry transformations generated by $Q$,
i.e for the vacuum
\begin{align}
  e^{i\epsilon Q}\mid 0\rangle =\mid 0 \rangle  \rightarrow Q \mid 0 \rangle=0
\end{align}
where $Q$ is a generator of the symmetry group under consideration. The vacuum
$\mid 0\rangle$ is unique and defined as the state of the system which 
\begin{align}
  \langle 0\mid H\mid 0\rangle =\mathrm{min}.
\end{align}
So the generators $Q$ generate a symmetry i.e. when
$[Q,H]=0$, and $\mid a\rangle$ and $\mid a'\rangle$ belong to the same
multiplet then $H\mid a\rangle=E_a\mid a\rangle$ implies that $H\mid
a'\rangle=E_a\mid a'\rangle$. This means that $a$ and $a'$ are degenerate states.

In the Goldstone mode the Lagrangian is invariant under the symmetry
transformations generated by $Q$, but $Q\mid 0\rangle \neq 0$. The consequence
of this will be discussed in detail in the
next chapters.

\subsubsection{Spontaneous symmetry breaking}

In the following we consider an easy example that describes the theory behind 
spontaneous symmetry breaking. This happens when the symmetry of the 
Lagrangian is not shared by the groundstate. We will develop this idea and
consider the linear
sigma- and the non-linear sigma-model, where an effective Lagrangian is derived
from the symmetry properties of the theory. Later on we use the same symmetry
arguments for the QCD and QFD Lagrangians to determine their form.

Start by consider a Lagrangian with a discrete reflection symmetry, $\phi
\rightarrow -\phi$, i.e the $\phi^4$ theory Lagrangian
\begin{align}
  \label{eqss1}
  \mathcal{L}=
  \frac{1}{2}(\partial_\mu \phi)^2 - 
  \frac{1}{2} \mu^2 \phi^2 -
  \frac{\lambda}{4!} \phi^4
\end{align}
where $\phi$ is a spin-zero scalar field, $\lambda$ and $\mu$  are constants. 
The field $\phi_0$ is the classical minimum of the potential
\begin{align}
  V(\phi)=\frac{1}{2} \mu^2 \phi^2+ \frac{\lambda}{4!} \phi^4
\end{align}
which has two minima for $\mu^2<0$, 
\begin{align}
  \phi_0=\pm v=\pm \sqrt{- \frac{6\mu^2}{\lambda}}.
\end{align}
The constant $v$ is the vacuum expectation value of $\phi$. Now expand around one
of the two minima,
\begin{align}
  \phi(x)=v+\sigma(x)
\end{align}
and rewrite $\mathcal{L}$ in terms of $v$ and $\sigma$. Dropping the constant
terms, the following $\mathcal{L}$ is obtained
\begin{align}
  \mathcal{L}=
  \frac{1}{2}(\partial_\mu \sigma)^2 - 
  \frac{1}{2}(-2 \mu^2 )\sigma^2 -
  \sqrt{- \frac{\lambda \mu^2}{4!}} \sigma^3 - 
  \frac{\lambda}{4!} \sigma^4
\end{align}
which describes a simple scalar field of mass $\sqrt{-2 \mu^2}$, with $\sigma^3$
and $\sigma^4$ interactions. The reflection symmetry in Eq.~\eqref{eqss1} is no
longer present and you say that it is hidden inside the Lagrangian.

\subsubsection{The linear sigma model}

A generalization of the discrete symmetry breaking above is the
linear sigma model, which has a broken continuous symmetry. 
This is a phenomenological model for light mesons at low energy. 
The model consist out of N fields and has a $\phi^4$ interaction 
that is invariant under rotation of the N fields. The Lagrangian is
\begin{align}
  \mathcal{L}=
  \frac{1}{2}(\partial_\mu \phi^i)^2 - 
  \frac{1}{2}\mu^2(\phi^i)^2-
  \frac{\lambda}{4}((\phi^i)^2)^2\label{eql1}
\end{align}
where i=1, ..., N and is summed over. 
The last two terms form the potential and are often written as
\begin{align}
  \label{eql5}
  V(\phi^i)=\frac{1}{2} \mu^2(\phi^i)^2+\frac{\lambda}{4}((\phi^i)^2)^2.
\end{align}
\begin{figure}[t]
  \includegraphics[angle=270, scale=0.32]{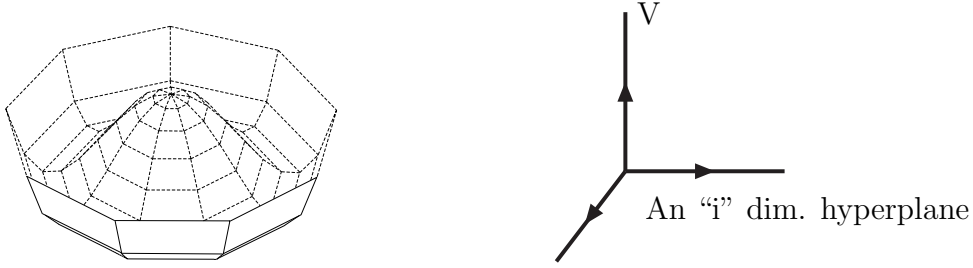}
  \hspace{1.6cm}
    \begin{picture}
      (60,0)(300,90)
      \SetWidth{1.5}
      
      \ArrowLine(300,0)(300,60)
      \ArrowLine(300,0)(360,0)
      \ArrowLine(300,0)(274,-34)

      \Text(310,60)[]{V}
      \Text(370,-15)[]{An ``i'' dim. hyperplane}
    \end{picture}
    \caption{Shows the ``Mexican hat potential'' in Eq.~\eqref{eql5} where the 
      potential energy $V$ is a function of $\phi^i$ and the plane parallel to $V$ is 
      thought of as an "i" dimensional hyperplane.}
\end{figure}
%
%
%
The Lagrangian in Eq.~\eqref{eql1} is invariant under the symmetry transformation
\begin{align}
  \phi^i \rightarrow R^{ij} \phi^j
\end{align}
for an $N\times N$ orthogonal matrix R which imply that $\mathcal{L}$ has an $O(N)$
symmetry. The potential $V$ has a local maximum at
$\phi^i=0$ for $\mu^2<0$. The minimum (ground state) can be found through
$\frac{\partial V}{\partial \phi^i}$ and thought of as a
multidimensional circle. Excitations in the N:th direction (radial direction)
require a shift away from the minimum, whilst an excitation along the circle,
where the potential is flat, corresponds to the other $N-1$ directions. Here an
intuitive physical picture can be seen, and as will be seen later, the excitation
in the radial direction corresponds to a massive particle and an excitation
along the flat potential gives rise to massless particles. The chosen minimum
of the potential, $\phi_0$, can be written in the following way:
\begin{align}
  \phi_0^i(x)&=0          && i=1, ..., N-1
  \nonumber\\
  \phi_0^i(x)&=v          && i=N
\end{align}
When a perturbation expansion is performed, the fluctuations around $\phi_0$ have to 
be considered instead of the fluctuations around $\phi=0$. This means that
\begin{align}
  &\phi^i(x)=(\varphi^k(x),v+ \tilde{\sigma}(x))   && k=1,..., N-1 \label{eql2}
\end{align}
where $\varphi^k(x)$ are the light meson fields and $v$ the vacuum expectation value.
The original $O(N)$ symmetry is then hidden (broken) and the remaining symmetry is 
$O(N-1)$, which rotate the $\varphi^k(x)$ among themselves. Now Eq.~\eqref{eql1} can 
be rewritten by replacing the old $\phi^i$ with the new one in Eq.~\eqref{eql2}. The 
following result is then obtained
\begin{eqnarray}
  \label{eql3}
  \mathcal{L} &=&
  \frac{1}{2}(\partial_\mu \vec{\varphi}^k)^2 +
  \frac{1}{2}(\partial_\mu
  \tilde{\sigma})^2 -
  \frac{1}{2}(-2\mu^2) \tilde{\sigma}^2 - 
  \sqrt{-\lambda \mu^2} \tilde{\sigma}^3 - 
  \sqrt{-\lambda \mu^2} (\varphi^k)^2 \tilde{\sigma} 
  \nonumber\\ && -
  \frac{\lambda}{4} \tilde{\sigma}^4 -
  \frac{\lambda}{2}(\varphi^k)^2 \tilde{\sigma}^2 -
  \frac{\lambda}{4}((\varphi^k)^2)^2
\end{eqnarray}
where $N-1$ massless meson fields and one massive sigma field have appeared.

If N=4
\footnote{N=4 is useful because $O(4) \cong SU(2) \times SU(2)$ which
  means that the exponential representation with the Pauli matrices can be
  used.} 
an $O(4)$ symmetry is achieved, which means that 
\begin{equation}
  \mathbf{\phi}=\left(
    \begin{array}
      {c}\sigma\\\pi_1\\\pi_2\\\pi_3
    \end{array}
  \right)
  \rightarrow
  \left(
    \begin{array}
      {c}\sigma '\\\pi_1 '\\\pi_2 '\\\pi_3 '
    \end{array}
    \right)
    =
    O
    \left(
    \begin{array}
      {c}\sigma\\\pi_1\\\pi_2\\\pi_3
    \end{array}
    \right)
\end{equation}
where $O$ is an orthogonal $4\times 4$ matrix that rotate the matrix
$\phi$. This means that the four primed
fields are linear combinations of the unprimed fields, and that the
fields transform linearly. If $\sigma=v+\tilde{\sigma}$ is considered, where
the sigma field is a scalar field and has a non-vanishing vacuum expectation value, 
then Eq.~\eqref{eql3} can be rewritten as
\begin{align}
  \mathcal{L}=
  \frac{1}{2} \partial_\mu \vec{\varphi} \cdot \partial^\mu \vec{\varphi} +
  \frac{1}{2} \partial_\mu \sigma \partial^\mu \sigma - 
  \frac{\mu^2}{2}(\sigma^2+\vec{\varphi}^2) - 
  \frac{\lambda}{4}(\sigma^2+\vec{\varphi}^2)^2
\end{align}
Using a representation called the exponential representation
\begin{equation}
  S\equiv \sqrt{\sigma^2+\vec{\varphi}^2}-v
  \,,\quad
  U\equiv \exp(i\vec{\tau} \cdot \vec{\varphi '}/v)\,.
\end{equation}
The resulting Lagrangian can be written in the form
\begin{align}
  \label{eql4}
  \mathcal{L}=\frac{1}{2} ((\partial_\mu S)^2-2\mu^2
  S^2)-\lambda vS^3-\frac{\lambda}{4} S^4 + \frac{(v+S)^2}{4} \langle
  \partial_\mu U \partial^\mu U^\dagger \rangle
\end{align}
where $\langle...\rangle$ is the trace in flavor space. Since we have done the
switch $\phi \rightarrow U$, using local group
arguments\footnote{$\vec{\varphi}'$ is determined from $\sigma +i\vec{\tau} \cdot
  \vec{\varphi}=(S+v)U$.}, the 
linear transformation is not valid anymore. The matrix $U$ is invariant under 
$SU(2)_L \times SU(2)_R$, but this is a non-linear transformation, which can
be seen explicitly when we Taylor expand
\begin{equation}
  U\simeq 1+\frac{1}{v}\vec{\tau} \cdot \vec{\varphi}'-\frac{(\vec{\tau} \cdot
  \vec{\varphi}')^2}{v^2}+\cdots
\end{equation}
and use that $U\rightarrow g_LUg_R^\dagger \Rightarrow \vec{\tau}\cdot
\vec{\varphi}'\rightarrow g_L\vec{\tau} \cdot \vec{\varphi}'g_R^\dagger+\cdots$.
So the Lagrangian in Eq.~\eqref{eql4} can be thought of as the non-linear 
representation of the linear sigma model.

\subsubsection{The non-linear sigma-model}

When physics is studied at a specific energy scale only particles that can be 
created inside this energy scale have to be considered. The heavy
fields do then not turn up explicitly, but can be seen through virtual
effects. When an effective $\mathcal{L}$ (see chapter 2.5) is used at low energies 
the heavy 
fields do not have to be included, but the virtual effects are included via
LECs connected to only light fields. There is a theorem called the
decoupling theorem that says that 
all effects from heavy fields will reveal themselves through renormalization of 
coupling constants or they can be ignored because of the heavy mass
suppression. At low energy, Eq.~\eqref{eql4} can be expanded to a very
interesting form. By only considering lowest order $\varphi '\varphi '$-interactions 
the only 
interesting term is the one without the field S, so all diagrams with S
interactions and S propagators are neglected. The non-linear sigma model is 
achieved when S $\rightarrow$ 0, which implies that 
\begin{align}
  \mathcal{L}= \frac{v^2}{4} \langle
  \partial_\mu U \partial^\mu U^\dagger \rangle\label{eqnl1}.
\end{align}
This is of great importance as we will see later on. Without external fields
and EM interactions this term is the only term of $\mathcal{O}(p^2)$.
 
 



\subsubsection{Goldstone bosons}

That massless particles show up when a continuous symmetry is broken, is a
consequence from Goldstone's theorem\cite{gb1}. The theorem says that that for every 
spontaneous broken symmetry, the theory must contain a massless particle. So
the result from an $O(N)$-symmetric theory as for the linear sigma model with
$N-1$ broken symmetries, is $N-1$ massless particles. The massless particles 
are called Goldstone bosons (GB).

As I mentioned before the Lagrangian is invariant, but $Q\mid{0}\rangle \neq
0$ for a number of generators. This implies that in the vacuum
there are operators that form states of particles, denoted $\mid{\pi^a(k)}
\rangle$. This fact will be used later on when we show that the 
non-linear sigma model is equivalent with $\mathcal{L}_{QCD}$.

In the earlier chapters the symmetry of the QCD Lagrangian was
discussed and in the chiral limit a $SU(3)_L\times{SU(3)_R}\times
{U(1)_A}\times{U(1)_V}$ symmetry was achieved.
If only the two lightest quarks had been considered then the QCD
Lagrangian would have a $SU(2)_L\times SU(2)_R\times
U(1)_A\times U(1)_V$. The first two groups corresponds
to $SO(4)$ (true under Lie algebra). The interesting part is that since 
$SO(4)\rightarrow SO(3)$ under
spontaneous symmetry breaking (where $SO(3)$ corresponds to $SU(2)$ under simular 
group argument as above)\footnote{$SU(2)$ is locally identical
  to the rotations in three dimensions, which form the group $SO(3)$, but they
  differ globally i.e. for rotations that are not infinitesimally small. The two groups 
  have the same Lie algebra, i.e. their generators have the same commutator
  relations} $SU(2)_L\times SU(2)_R\rightarrow SU(2)_{V=L+R}$, 
where ${SU(2)_{V=L+R}}$ is a subgroup of the chiral symmetry. 
Using the same arguments as above $G\equiv SU(3)_L\times SU(3)_R\rightarrow SU(3)_
{V=L+R}\equiv H$. Since the global axial part of $\mathcal{L}_{QCD}$ in the 
chiral limit is broken and the fact that we are working in $SU(3)$, eight
 \footnote{The number of GB is determined by the symmetry
  properties of the groups. The $G$ group denote the symmetry group of the
  Lagrangian, with $n_G$ generators. The $H$ subgroup, that leaves the vacuum
  after spontaneous symmetry breaking invariant, has $n_H$ generators. Every
  generator that does not annihilate the vacuum creates one massless GB, i.e. the
  total number of GB is $n_G-n_H$. In our case $n_G=16$ and $n_H=8$, which means that 
  eight Goldstone bosons appear.}  
pseudoscalar Goldstone bosons (GB) appear as a consequence of Goldstone's
theorem.

The massless particles showing up form the light meson octet containing:
$\pi^{\pm}$, $\pi^0$, $K^{\pm}$, $K^0$, $\bar{K^0}$, $\eta$. The reason for
this is that the light particles have the correct quantum numbers to be 
associated with the generators of the coset space $G/H$, which
parameterize the GB. Since the GB are massless in this theory, their masses 
have to be introduced in a different way. It can be added to
$\mathcal{L}_{QCD}$, but this will break the perfect chiral symmetries. The
adding of the mass terms to the QCD Lagrangian is analogous to adding a $a\Phi$
term to the potential in Eq.~\eqref{eql5}. Using a perturbative ansatz, one obtains
massive GB, where the squared masses are proportional to the
breaking parameter a. Since the masses of the three light quarks are small 
compared to the breaking scale of chiral symmetries ($\sim1
GeV$)\footnote{Around this energy the spin-one states ($\rho, K^*, \phi$) can be
  created with the same quark
  content as the light meson octet. The spin-one particles can not be
  GB, since GB have spin zero if Lorentz invariance is a good
  symmetry.}, the
explicit adding of the mass  can be treated as a small
perturbation. The fact that the pseudoscalar-meson masses are much
smaller than the typical hadronic scale reinforces this assumption. 

One more interesting feature of this is that the non-vanishing masses of the 
light pseudoscalars in the "real" world, are related to the explicit symmetry 
breaking in QCD due to the light quark masses which are believed to acquire
their mass from the Higgs mechanism.

\subsection{Chiral perturbation theory}

Chiral perturbation theory (ChPT) is based upon the following two basic 
assumptions:
\begin{enumerate}
\item
  The spontaneous breakdown of the $SU(3)_L\times SU(3)_R$ symmetry to
  $SU(3)_V$ for the QCD Lagrangian in the chiral limit.
\item
  Close to the chiral limit the mass terms of the light quarks can be treated as small
  perturbations.
\end{enumerate}

It is very useful for strong interactions in the low energy sector. 
As seen in previous chapters the symmetry of the QCD Lagrangian is
spontaneously broken to $SU(3)_V\times{U(1)_V}$ which introduces eight massless
Goldstone bosons in the form of the ($\pi$, $K$, $\eta$) octet. Since $U(1)_V$ 
corresponds to the baryon number it plays no further role in low energy meson
physics.

When the quark masses are added to $\mathcal{L}_{QCD}$, the pseudoscalar mesons
aquire mass through an explicit chiral symmetry breaking. The mass is added
through a chiral invariant coupling of the mesons to the scalar and
pseudoscalar external currents (see appendix C) s and p, respectively, 
\begin{align}
  \chi=2B_0(s+ip)\rightarrow 2B_0 g_R(s+ip)g_L^\dagger
\end{align}
where $g_{L, R}{\in}SU(3)_{L,R}$, $B_0$ is related to the quark condensate
$\langle0|\bar{u}u|0\rangle=-F_0^2 B_0[1+O(m_ q)]$, p is a 3$\times$3 matrix 
containing eight pseudoscalar external fields related to the $\bar{q}
\gamma_5 q$ current. The scalar external currents includes the masses for the 
three light quarks, 
\begin{align}\label{eqchp1}
  s=\mathcal{M}+\cdots=\left( 
    \begin{array}{ccc}
      m_u &     &    \\
      & m_d &    \\
      &     & m_s
    \end{array} \right)
  +\cdots
\end{align}
and are related to the $\bar{q} q$ current.
To get rid of the part of the external current s which is of no importance to
us, put it exactly equal to the mass matrix. By doing this 
$\chi \rightarrow \hskip-1.2em/$ $ g_L \chi g_R^\dagger$,
so the chiral symmetry breaks, 
but the quark masses are small compared to the constituent mass, so this 
can be treated as a perturbation. And since the creation of the GB is based upon 
the $SU(3)_L\times SU(3)_R$ symmetry which is broken by the perturbative mass, the GB
change to pseudo-GB. 

ChPT is a non-renormalizable theory, because of the ultraviolet divergences
produced by the loops. For a non-renormalizable theory infinite numbers of
counterterms have to be added to make it finite. The problem can be avoided by
using effective quantum field theory (see effective Lagrangian chapter), which 
indicate that the energy
region of interest is below a certain cut-off parameter $\Lambda$, and then
the number of counterterms needed at each level of expansion is finite. At 
increasing expansion powers, the number of needed counterterms increases 
drastically.
At very low energies, the contributions from higher orders are small, that is 
why only terms up to $\mathcal{O}(p^4)$ are considered here, see
Eq.\eqref{eqi1}. The finite parts for each power contains a finite coupling 
constant and it is some of those constants that are going to be determined 
in this thesis. The problem is that they cannot be calculated directly from 
the QCD and QFD Lagrangians. 

\subsection{Effective Lagrangians}

The reason why an effective Lagrangian is used is that at high energies QCD is 
perturbative, i.e it can be expanded in terms of $\alpha_s$, which is small 
for large energies. At low energies the theory is highly non-perturbative since
$\alpha_s$ is large and we therefore need to use effective Lagrangian
techniques. When effective Lagrangians are used
instead of the original Lagrangians, they correspond to the same symmetries
but generate different Feynman diagrams.  

Most theories in physics are an approximation to the real world. The
perturbative treatment is a common and well known method. For the task of
this thesis a perturbative theory is needed. The problem can be solved if a
light pseudoscalar meson representation is used. The representation is not
expanded in terms of $\alpha_s$, but instead in terms of quark masses and 
momenta. This is known as chiral expansion. 

For low enough energies only a few relevant degrees of freedom are needed to 
describe
the theory. The non-relevant d.o.f can be integrated out and encoded into the
LECs, just as for the non-linear
sigma model. The effective low energy action that is described
by a number of relevant and non-relevant fields, can be written in the form
of external fields in the GB low energy theory. The earlier mentioned
Lagrangian is then modified, and the external fields can be included in a
chiral invariant way. 


\subsubsection{Power counting}

Power counting is a very useful way to organize the powers in the chiral
expansion. The ordering in the context of Feynman rules is that, derivatives
generate four momenta and the quark mass terms have the same dimension as two
derivatives. So when we are on-shell $m_{meson}^2 = p^2$, but when we are off-shell
$e^2$ has the same counting as the four momenta squared. The external photon
fields are of order p since they occur together with a momentum in the
covariant derivative. An illustrative example can be seen below:
\begin{center}
  \begin{picture}
    (300,200)(0,-200)
    \SetWidth{1.}

    \Line(0,0)(50,-50)
    \Line(50,0)(0,-50)
    \put(25,-25){\circle*{3}}
    \Text(120,-25)[]{meson vertex}
    \Text(220,-25)[]{$\sim p^2$}

    \Line(0,0)(50,-50)
    \Line(50,0)(0,-50)
    \put(25,-25){\circle*{3}}
    \Text(120,-25)[]{meson vertex}
    \Text(220,-25)[]{$\sim p^2$}

    \Line(0,-75)(50,-75)
    \Text(120,-75)[]{meson propagator}
    \Text(220,-75)[]{$\sim 1/p^2$}

    \Text(20,-125)[]{$\int d^4 p$}
    \Text(120,-125)[]{loop integral}
    \Text(220,-125)[]{$\sim p^4$}

    \Line(0,-150)(12,-175)
    \Line(0,-200)(12,-175)
    \Line(50,-150)(37,-175)
    \Line(50,-200)(37,-175)
    \CArc(25,-175)(12.5,0,360)
    \GCirc(12,-175){1.5}{0}
    \GCirc(37,-175){1.5}{0}
    \Text(150,-175)[]{$\sim (p^2)^2(1/p^2)^2p^4=p^4$}
  \end{picture}
\end{center}
In the figure above, the loop diagram is of  $\mathcal{O}(p^4)$ and is 
generated by tree level diagrams from $\mathcal{L}_2$. The loop integrals
generate divergences which can be canceled by adding tree diagrams, from the
$\mathcal{L}_4$ terms. The counterterms needed to cancel the divergences 
created by the loop diagrams of $\mathcal{O}(p^n)$ comes from the LECs of 
the same order, which implies two things. First the good thing which 
is that the loop integral divergences are canceled order by order. The other thing 
is that the number of coefficients needed to describe the
theory increases dramatically for increasing orders, which makes an expansion to 
higher orders almost impossible.

For $\pi \pi$ scattering in pure QCD, an infinite number of
diagrams have to be considered since they are equally important. If instead an
effective theory is considered, the diagrams can be divided into groups of
different powers of $p^2$. For low energy, only few powers of $p^2$ have to be
considered. The physical picture can then be described by a small number of diagrams, 
which is a good approximation of QCD for low energy.

\subsubsection{QCD at low energy}

At low energies quarks do not behave like free quarks, instead they behave like
asymptotic light meson fields. The effective theory of QCD is formulated in
this way. It can be expanded in a series of infinite numbers of derivative
terms, i.e
\begin{equation}
  \mathcal{L}_{eff}=\mathcal{L}_2+\mathcal{L}_4+\mathcal{L}_6+\cdots
\end{equation}
From the properties of the GB and the symmetry of QCD
\footnote{With this I mean by adding local contributions to the external
  fields the transformations associated with the $J$ symmetry in Eq.~\eqref{eqs2} 
  implies certain invariances, which are used to derive the structure of the
  effective Lagrangians. All this follows from Gasser \& Leutwyler \cite{ntloel1}.}
, the easiest form of the leading order terms is
\begin{equation}
  \label{eqqcdale1}
  \mathcal{L}_2^{QCD}=
  \frac{F_0^2}{4}\langle \partial ^\mu U^\dagger \partial _\mu U+
  2B_0 \mathcal M (U^\dagger+U) \rangle
\end{equation}
where $\mathcal{M}$ is the mass matrix in Eq.~\eqref{eqchp1}. U contain the eight 
pseudoscalar meson fields and has the following properties:
\begin{align}
  \label{eqloel2}
  &UU^+=\boldsymbol{1},        && \det U=1    \nonumber\\
  &U=\exp{(\frac{i\Phi}{F_0})},   && \Phi=\sum_{a=1}^{8}{\lambda_a}{\varphi_a},
\end{align}
where
\begin{align}
  \label{eqloel3}
  \Phi =
  \sqrt{2}\left( 
    \begin{array}{ccc}
      \frac{\pi^0}{\sqrt{2}}+\frac{\eta_8}{\sqrt{6}} & {\pi^+} & K^+\\
      \pi^- & -\frac{\pi^0}{\sqrt{2}}+\frac{\eta_8}{\sqrt{6}} & K^0\\
      K^- & \bar{K}^0 & -\frac{2\eta_8}{\sqrt{6}}
    \end{array} \right).
\end{align}
The observant reader can see that in the chiral limit, the vacuum 
expectation value $v$ in the non-linear sigma model in Eq.~\eqref{eqnl1}
can be put equal to the pion decay coupling $F_0$. This coupling constant is 
defined from the differentiation of the lowest order classical action 
$S_ 2=\int{d^4x \mathcal{L_2}}$ with respect to external axial fields in the 
chiral limit, i.e.
\begin{equation}
  \langle 0\mid \frac{\delta S_2}{\delta a_\mu} \mid \pi^+(p) \rangle =
  \langle 0\mid \bar d \gamma^{\mu} \gamma_5 u \mid \pi^+(p)\rangle =
  i\sqrt2 F_0 p^\mu.
\end{equation}
We compare this with the pion decay constant which is derived as follows:
The decay of a $\pi^+$ into two leptons via a weak interaction can be
described by the tree level diagram
\begin{center}
  \begin{picture}
    (140,40)(0,0)
    \SetWidth{1.}

    \Line(5,5)(30,20)
    \Line(5,35)(30,20)
    \put(30,20){\circle*{3}}
    \Text(15,35)[]{u}
    \Text(15,5)[]{$\bar{d}$}

    \DashLine(30,20)(110,20){3}
    \Text(70,25)[]{$W^+$}

    \Line(110,20)(135,35)
    \Line(110,20)(135,5)
    \put(110,20){\circle*{3}}
    \Text(120,35)[]{$\mu^+$}
    \Text(120,5)[]{$\nu_\mu$}   
  \end{picture}  
\end{center}
Written in terms of matrix elements , this corresponds to 
\begin{equation}
 \langle W^+\mid \mathcal{L}\mid \pi^+(p)\rangle \sim
 \langle W^+\mid \bar W_\mu^+ \mid 0\rangle
\langle 0\mid \bar d \gamma^{\mu} \gamma_5 u \mid
  \pi^+(p)\rangle
\end{equation}
The pion decay constant is then defined to be the second term on the right
side. When the amplitude is calculated the following is achieved
\begin{equation}
  i\sqrt2 F_\pi p^\mu = \langle 0\mid \bar d \gamma^{\mu} \gamma_5 u \mid
  \pi^+(p)\rangle
\end{equation}
and the identification $F_\pi=F_0$ can be made.

\subsubsection{QFD at low energies}

The electromagnetic virtual interactions of order $e^2$ between pseudo-GB
to the leading order in the chiral expansion which was first
derived by R. Dashen \cite{qfdale1} and is described by the effective
Lagrangian 
\begin{equation}
  \label{eqQFD1}
  \mathcal{L}_2^{EM}=e^2 C_1 \langle {Q^2} \rangle
  +e^2C \langle QUQU^\dagger \rangle
\end{equation}
which is created from the symmetry properties for an effective low 
energy QFD theory. To ensure the chiral 
$SU(3)_L \times SU(3)_R$ symmetry in 
Eq.~\eqref{eqQFD1}, the local spurions\footnote{Spurions are dummy fields that are
  added to the Lagrangian to make it a singlet under the chiral group.} 
\begin{align}
  &Q_L \rightarrow g_L Q g_L ^\dagger
  \nonumber\\
  &Q_R \rightarrow g_R Q g_R ^\dagger
\end{align}
are introduced instead of the charge matrix Q. The first term in
Eq.~\eqref{eqQFD1} is a constant and will not be considered. So the resulting
Lagrangian is
\begin{equation}
  \label{eqQFD2}
  \mathcal{L}_2^{EM}=e^2C \langle Q_L U Q_R U^\dagger \rangle.
\end{equation}
The two spurions, $Q_L$ and $Q_R$, are different matrices. Since the
photon couples identically to the left respectively right, it can not
distinguish between the two Q matrices and for that reason we put them equal in our
effective Lagrangians.

\subsubsection{Leading order effective Lagrangian}

The effective Lagrangian of QCD including EM effects to leading order is in
the mesonic sector
\begin{align}
  \label{eqloel1}
  \mathcal{L}_2=
  -\frac{1}{4}F_{\mu\nu}F^{\mu\nu}-\frac{1}{2(1-\xi)}({\partial_\mu}{A^\mu})^2+
  \frac{1}{4}{F_0^2}\langle{D^\mu}{U^+}{D_\mu}U+
  {\chi}{U^+}+{\chi^+U}\rangle+e^2C\langle QUQ U^\dagger \rangle
\end{align}
where, the two first terms describe the kinetic energy and the gauge fixing
term\footnote{The main reason why the gauge fixing term is introduced is to avoid 
singularities when the photon mass equals zero.} for the virtual photons. When one 
takes the inverse of the two terms it will result in 
the photon propagator that we later on use to calculate the photon loops. 
$F_{\mu\nu}={\partial_\mu}{A\nu}-{\partial_\nu}
{A_\mu}$ is the electromagnetic field strength tensor of the photon field 
$A_\mu$ and $\xi$ is a gauge fixing term. The covariant derivative in 
Eq.~\eqref{eqloel1} conserves the local gauge 
symmetries since it includes the couplings to the photon field
$A_\mu$, the external vector $v_\mu$ and axial-vector currents $a_\mu$,
\begin{equation}
  \label{eqloel4}
  D_ \mu U=
  \partial_\mu-i(v_\mu+a_\mu)U+iU(v_\mu-a_\mu)=
  \partial _\mu -i r_\mu U+iU l_\mu
\end{equation}
where $v_\mu=\frac{\lambda^a}{2}v_\mu^a=e^2QA_\mu+\tilde{v}_\mu$, and 
$a_{\mu}=\frac{\lambda^a}{2}a_\mu^a$ is a 3$\times$3 matrix
associated to the quark current $\bar{q} \gamma_\mu \gamma^5 \lambda^a q$. 
$\tilde{v}_ \mu$ is associated with $\bar{q} \gamma_\mu \lambda^a q$, but this
current is not of interest to us. 
The Q matrix in Eq.~\eqref{eqloel1} is the charge matrix for the u, d, and s quark, 
\begin{align}
  \label{eqloel5}
  Q = \frac{1}{3}\left( 
    \begin{array}{ccc}
      2 &    &    \\
      & -1 &    \\
      &    & -1
    \end{array} \right).
\end{align}

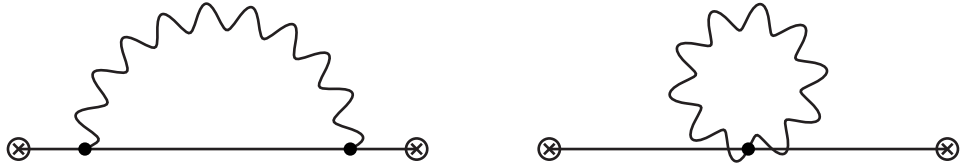
\begin{figure}[t]
  \begin{center}
    \begin{picture}
      (360,160)(-10,-10)
      \SetWidth{1.}

      \Line(0,0)(150,0)
      \PhotonArc(75,0)(50,0, 180){5}{10}

      \CArc(0,0)(4,0,360)
      \Line(2,2)(-2,-2)
      \Line(2,-2)(-2,2)
      \CArc(150,0)(4,0,360)
      \Line(148,-2)(152,2)
      \Line(152,-2)(148,2)

      \GCirc(25,0){2}{0}
      \GCirc(125,0){2}{0}

      \Line(200,0)(350,0)
      \PhotonArc(275,25)(25,0,360){5}{10}

      \CArc(200,0)(4,0,360)
      \Line(202,2)(198,-2)
      \Line(202,-2)(198,2)
      \CArc(350,0)(4,0,360)
      \Line(348,-2)(352,2)
      \Line(352,-2)(348,2)

      \GCirc(275,0){2}{0}
    \end{picture}
  \end{center}
  \caption{The two first types of one-loop contributions from
    Eq.~\eqref{eqloel1}. The crosses are external pseudoscalar currents and the 
    wiggly line is the virtual photon. The full lines are pseudoscalars.}
  \label{fig:ntloel1}
\end{figure}

As we have seen above, the lowest order Lagrangian, $\mathcal{L}_2$, is given
by the non-linear sigma model coupled to external fields. When calculating a 
process using only tree level
diagrams of $\mathcal{L}_2$, one represents the results of current
algebra. The tree level diagrams can be put together to one-loops diagrams
where we have an emission and absorption of a photon in the pseudo-GB theory, 
generated by the two-point function (Fig.~\ref{fig:ntloel1}), which include the 
vertices $\gamma P^+P^-$ and $\gamma \gamma P^+P^-$ incorporated in 
Eq.~\eqref{eqloel1}.
A calculation of one-loop diagrams with $\mathcal{L}_2$ vertices, leads to
divergences as I mentioned before. Due to Weinberg's counting these infinities
are of $\mathcal{O}(p^4)$ and can be canceled through renormalization by next-to-
leading order Lagrangian at $\mathcal{O}(p^4)$.

\subsubsection{Next-to-leading order effective Lagrangian}

The Lagrangian in Eq.~\eqref{eqloel1} is only the first term in the infinite chiral 
expansion series. Here we introduce the second term in the expansion. The
first part of it is of order $p^4$ and for the three light quark approximation 
in the pure strong sector ($Q=0$), it was 
first written down by Gasser and Leutwyler \cite{ntloel1}, while the EM part
of order $p^2e^2$ was first derived by Urech \cite{ntloel2}. 

The $\mathcal{L}_4$ Lagrangian contains all allowed operators 
that transform linearly and are invariant under the correct symmetries of QCD
and QFD. It is given by:
\begin{align}
  \label{eqntloel1}
  \mathcal{L}_4= \mathcal{L}^{QCD}_4+\mathcal{L}^{EM}_4=
  &L_1\langle D_\mu U D^\mu U^\dagger \rangle^2 +
  L_2\langle D_\mu U D^\nu U^\dagger \rangle^2 +
  L_3\langle D_\mu U D^\mu U^\dagger D_\nu U D^\nu U^\dagger\rangle
  \nonumber\\
  +&L_4\langle D_\mu U D^\mu U^\dagger \rangle 
  \langle \chi U^\dagger+U\chi^\dagger \rangle +
  L_5\langle (D_\mu U D^\mu U^\dagger)(\chi U^\dagger+U\chi^\dagger) \rangle
  \nonumber\\
  +&L_6\langle \chi U^\dagger+U\chi^\dagger \rangle^2 +
  L_7\langle \chi U^\dagger-U\chi^\dagger \rangle^2 +
  L_8\langle \chi U^\dagger \chi U^\dagger+U\chi^\dagger U\chi^\dagger \rangle
  \nonumber\\
  +&iL_9\langle R_{\mu \nu} D^\mu UD^\nu U^\dagger+
  L_{\mu \nu} D^\mu U^\dagger D^\nu U \rangle +
  L_{10}\langle R_{\mu \nu} U L^{\mu \nu} U^\dagger \rangle
  \nonumber\\
  +&H_1\langle R_{\mu \nu}R^{\mu \nu}+L_{\mu \nu}L^{\mu \nu} \rangle +
  H_2\langle \chi^\dagger \chi \rangle
  \nonumber\\
  +&e^2F_0^2K_1\langle D^\mu U^\dagger D_\mu U \rangle \langle Q^2\rangle +
  e^2F_0^2K_2\langle D^\mu U^\dagger D_\mu U \rangle \langle QUQU^\dagger \rangle
  \nonumber\\
  +&e^2F_0^2K_3\langle QU^\dagger D^\mu UQ D_\mu U^\dagger U+
  QU D^\mu U^\dagger Q D_\mu UU^\dagger \rangle
  \nonumber\\
  +&e^2F_0^2K_4\langle QU^\dagger D^\mu U Q D_\mu UU^\dagger \rangle
  \nonumber\\
  +&e^2F_0^2K_5\langle (D^\mu U D_\mu U^\dagger+
  D^\mu U^\dagger D_\mu U)Q^2\rangle
  \nonumber\\
  +&e^2F_0^2K_6\langle D^\mu U^\dagger D_\mu U QU^\dagger QU+
  D^\mu U  D_\mu U^\dagger QUQU^\dagger \rangle 
  \nonumber\\
  +&e^2F_0^2K_7\langle \chi ^\dagger U+U^\dagger \chi \rangle \langle Q^2\rangle
  +e^2F_0^2K_8\langle \chi ^\dagger U+U^\dagger \chi \rangle 
  \langle QUQU^\dagger \rangle
  \nonumber\\
  +&e^2F_0^2K_9\langle (\chi ^\dagger U+U^\dagger \chi+
  \chi U^\dagger+U \chi ^\dagger) Q^2\rangle
  \nonumber\\
  +&e^2F_0^2K_{10}\langle (\chi ^\dagger U+U^\dagger \chi) QU^\dagger QU+
  (\chi U^\dagger +U \chi ^\dagger) QU QU^\dagger \rangle
  \nonumber\\
  +&e^2F_0^2K_{11}\langle (\chi ^\dagger U-U^\dagger \chi) QU^\dagger QU+
  (\chi U^\dagger -U \chi ^\dagger) QU QU^\dagger \rangle
  \nonumber\\
  +&e^2F_0^2K_{12}\langle UD^\mu U^\dagger [\nabla ^R_{\mu}Q,Q]+
  U^\dagger D^\mu U [\nabla ^L_{\mu}Q,Q]\rangle
  \nonumber\\
  +&e^2F_0^2K_{13}\langle \nabla ^R_{\mu}QU \nabla _L^{\mu}QU^\dagger \rangle +
  e^2F_0^2K_{14}\langle \nabla ^R_{\mu}Q\nabla _R^{\mu}Q+
  \nabla ^L_{\mu}Q\nabla _L^{\mu}Q\rangle
\end{align}
where the covariant derivatives $\nabla _{\mu}^{L(R)}$ are defined as
\begin{equation}
  \nabla _{\mu}^{L(R)}Q=\partial _\mu Q -i[v_\mu-(+)a_\mu,Q]
\end{equation}
and the $L$ and $R$ field strengths are
\begin{align}
  &L^{\mu \nu}=\partial ^\mu l^\nu -\partial ^\nu l^\mu -i[l^\mu,l^\nu]
  \nonumber\\
  &R^{\mu \nu}=\partial ^\mu r^\nu -\partial ^\nu r^\mu -i[r^\mu,r^\nu].
\end{align}
The coupling constants $L_i$, $H_i$ and $K_i$ are defined as 
\begin{align}
  &L_i=\Gamma_i\lambda + L_i^r(\mu),
  \nonumber\\
  &H_i=\Delta_i\lambda + H_i^r(\mu),
  \nonumber\\
  &K_i=\Sigma_i\lambda + K_i^r(\mu),
\end{align}
where the $\Gamma_i$ and $\Delta_i$ constants are determined by Gasser and
Leutwyler in \cite{i1} and the $\Sigma_i$ constants by Urech in \cite{ntloel2}.
$\lambda$ is the divergent part of the 
coupling constants and can be determined through dimensional regularization and 
$\overline{MS}$ regularization. The $L_i^r(\mu)$,
$H_i^r(\mu)$ and $K_i^r(\mu)$ are, respectively, finite parts, which we are
going to determine. These coefficients are renormalized and depend on the
arbitrary renormalization scale $\mu$.

\section{How to determine the EM coefficients}

\emph{The main idea behind the calculations of the $K_i$s is based on the 
investigation of a single two-point function using two different methods, 
and comparing them in the end. On the RHS we use pure ChPT to $\mathcal{O}(p^4)$, 
describing both QCD and QFD with internal photons. The amplitudes 
for the relevant Feynman diagrams on this side are generated by 
$\mathcal{L}_{eff}=\mathcal{L}_{2}+\mathcal{L}_{4}+\cdots$. 
On the other side, the LHS, we use external photons coupled to an effective Lagrangian 
with a QCD symmetry, 
$\mathcal{L}_{eff}^{QCD}=\mathcal{L}_{2}^{QCD}+\mathcal{L}_{4}^{QCD}+\cdots$.} 
\vspace{0.5cm}

First a summary of the plan:
\begin{itemize}
\item
  Investigate the Lagrangians $\mathcal{L}_2$ in Eq.~\eqref{eqloel1} and 
  $\mathcal{L}_4$ in Eq.~\eqref{eqntloel1}.
\item
  Expand the U matrix and plug in the matrices in Eq.~\eqref{eqloel3}, 
  Eq.~\eqref{eqloel4} and Eq.~\eqref{eqloel5}.
\item
  Consider the relevant Feynman diagrams.
\item
  Integrate over the momenta.
\item
  Calculate the amplitudes for the relevant processes.
\item
  Compare the different amplitudes generated by $\mathcal{L}_{eff}^{QCD}=
  \mathcal{L}_{2}^{QCD}+\mathcal{L}_{4}^{QCD}+\cdots$ on the left hand side (LHS)
  with $\mathcal{L}_{eff}=\mathcal{L}_{2}+\mathcal{L}_{4}+\cdots$ on the right
  hand side (RHS).
\end{itemize}

The main result in the chapters before is the Lagrangian in Eq.~\eqref{eqntloel1} of 
$\mathcal{O}(p^4)$ (where the $e^2p^2$ tree-level contribution from
Eq.~\eqref{eqntloel1} also is of this order). By expanding U in terms of the meson
matrix $\Phi$ and plugging in 
all the matrices into the Lagrangian, the terms that build up the Feynman 
diagrams are achieved. To handle all the terms by hand is a very bothersome
task, since there are so many of them. Instead we use the computer program
Form to calculate the terms. They are carefully investigated and
divided up into certain Feynman diagrams, e.g $\pi \pi$ interaction with
different kinds of loop diagrams. The terms from the Lagrangian are
translated according to the Feynman rules and the amplitude for each diagram is
calculated. This is easier said than done, since the diagrams with a loop
contain a loop integral over every relevant four-momenta. These 
integrals are calculated with different methods, depending if it is on the RHS 
or the LHS. The reason for this is that the two sides are related to different 
theories describing the same two-point function (appendix B).

We will now investigate the two sides separately.

\begin{description}
\item[Left hand side, (LHS)]
  
  Here we use a Lagrangian with a QCD symmetry including external
  virtual photons. There exist no divergences except the ones that can be absorbed
  with renormalization into $m_q$, $\alpha_s$ and $\alpha_{EM}$. The following
  picture illustrates the process
  \begin{center}
    \begin{picture}
      (80,80)(0,0)
      \SetWidth{1.}
      
      \GCirc(40,50){30}{0.9}
      \GCirc(5,50){5}{1}
      \Line(0,45)(10,55) 
      \Line(10,45)(0,55) 
      
      \GCirc(75,50){5}{1}
      \Line(70,45)(80,55)
      \Line(80,45)(70,55)
      
      \PhotonArc(40,70)(25,0,180){2}{10}
      \GCirc(64,69){2}{0}
      \GCirc(16,69){2}{0}
      
      \SetWidth{2.}
      
      \LongArrow(45,40)(70,10)
      \Text(85,10)[]{QCD}
    \end{picture}
  \end{center}
  where the figure describes a two-point function, with a virtual photon connected
  to it. The "bubble" generated by the Green function can be 
  thought of as a propagator describing a process from one side to another. 

  The vital point is that since we are dealing with virtual photons an
  integration over the whole four-momentum space has to be performed. This is
  going to give us divergences, which are absorbed as I mentioned before. The 
  integrals over the momentum are divided up into two parts with the help of a 
  cut-off (not the same cut-off used in the renormalization chapter) which 
  respectively is determined by a different method. The two parts are separated into 
  a long-distance (LD) and a short-distance (SD) contribution, with the help an
  euclidian cut-off $\Lambda$. The cut-off also works as a
  matching variable where a plateau in $\Lambda$ indicates matching. 
  After passing from Minkowski space to Euclidian space with the help of a Wick
  rotation\footnote{This means that the zeroth component of the four-momenta, 
    $p^0=ip^0_E$, where $p^0$ is in the Minkowski space and $p^0_E$ in the
    euclidian space}, the integrals can be written in the following form
  \begin{equation}
    \int_{0}^{\infty} dr_E= \int_{0}^{\Lambda} dr_E+\int_{\Lambda}^{\infty} dr_E.
  \end{equation}
  where $r_E$ is the momentum.
  The first integral, the LD part, is evaluated using ChPT to 
  $\mathcal{O}(p^4)$ which we will explain more in a moment. The second
  integral, called the SD part, is calculated with perturbative QCD. 
  
  \begin{itemize}
  \item[\bf{SD}]
  
    We evaluate the photons which have a momentum larger than 
    $\Lambda$. This is done through an expansion in powers of $1/\Lambda$, 
    where contributions up to next to leading order are considered. The essence of 
    the physical picture can be illustrated through a simple one-loop diagram.
    \begin{center}
      \begin{picture}
	(155,60)(0,0)
	\SetWidth{1.}
	
	\Line(5,-10)(40,20)
	\Line(5,50)(40,20)
	\put(40,20){\circle*{4}}
	
	\Photon(40,20)(120,20){3}{10}
	\Gluon(23,35)(137,35){3}{10}
	\Text(80,8)[]{$Q^2$}
	\put(137,35){\circle*{4}}
	\put(23,35){\circle*{4}}

	\Line(120,20)(155,50)
	\Line(120,20)(155,-10)
	\put(120,20){\circle*{4}}
      \end{picture}
    \end{center}
    \vspace{0.5cm}
    We use large photon momenta, so that the $Q^2$ is big. This means that a gluon 
    or a quark has to be included to run back with the momentum Q, otherwise the 
    diagram is heavily suppressed. The suppression comes from the fact that if only 
    ``one-way'' diagrams are considered, where the energy flows in one direction 
    then the 
    creation of the final hadron is suppressed since the quarks carry too much energy.

    In the calculations of the SD part we consider four types of different 
    contributions, each associated with one type of diagram and expanded to order 
    $1/\Lambda^2$.
    \begin{figure}[h]
      \begin{center}
	\begin{picture}
	  (360,60)(-10,-10)
	  \SetWidth{1.}
	  
	  \ArrowLine(0,0)(35,0)
	  \ArrowLine(35,0)(115,0)
	  \ArrowLine(115,0)(150,0)
	  
	  \PhotonArc(75,0)(40,0,180){5}{10}
	  
	  \GCirc(35,0){2}{0}
	  \GCirc(115,0){2}{0}
	  
	  \Text(75,-20)[]{(a)}
	  \ArrowLine(200,0)(235,0)
	  \ArrowLine(235,0)(275,0)
	  \ArrowLine(275,0)(315,0)
	  \ArrowLine(315,0)(350,0)
	  
	  \PhotonArc(275,0)(40,0,180){5}{10}
	  \Text(275,0)[]{X}
	  
	  \GCirc(235,0){2}{0}
	  \GCirc(315,0){2}{0}
	  \Text(275,-20)[]{(b)}
	\end{picture}
      \end{center}
      \begin{center}
	\begin{picture}
	  (360,100)(-10,-10)
	  \label{htdtemc1}
	  \SetWidth{1.}
	  
	  \ArrowLine(0,0)(35,0)
	  \ArrowLine(35,0)(115,0)
	  \ArrowLine(115,0)(150,0)
	  \ArrowLine(0,60)(35,60)
	  \ArrowLine(35,60)(115,60)
	  \ArrowLine(115,60)(150,60)
	  
	  \GCirc(35,0){2}{0}
	  \GCirc(115,0){2}{0}
	  \GCirc(35,60){2}{0}
	  \GCirc(115,60){2}{0}
	  
	  \Photon(35,60)(35,0){5}{7} 
	  \Gluon(115,60)(115,0){5}{7}
	  
	  \Text(75,-20)[]{(c)}
	  \Photon(235,60)(315,60){5}{7}
	  \ArrowLine(200,60)(235,60)
	  \ArrowLine(315,60)(350,60)
	  \ArrowLine(235,60)(275,30)
	  \ArrowLine(275,30)(315,60)
	  
	  \ArrowLine(200,0)(235,0)
	  \ArrowLine(235,0)(315,0)
	  \ArrowLine(315,0)(350,0)
	  
	  \Gluon(275,30)(275,0){5}{3}
	  \GCirc(275,30){2}{0}
	  \GCirc(275,0){2}{0}
	  
	  \GCirc(235,60){2}{0}
	  \GCirc(315,60){2}{0}
	  
	  \Text(275,-20)[]{(d)}
	\end{picture}
      \end{center}
      \caption{These are the four contributing diagrams. The curly line is the 
	gluon, the wiggly the photon, the plain line the quark and the cross 
	marks the external current.}
    \end{figure}
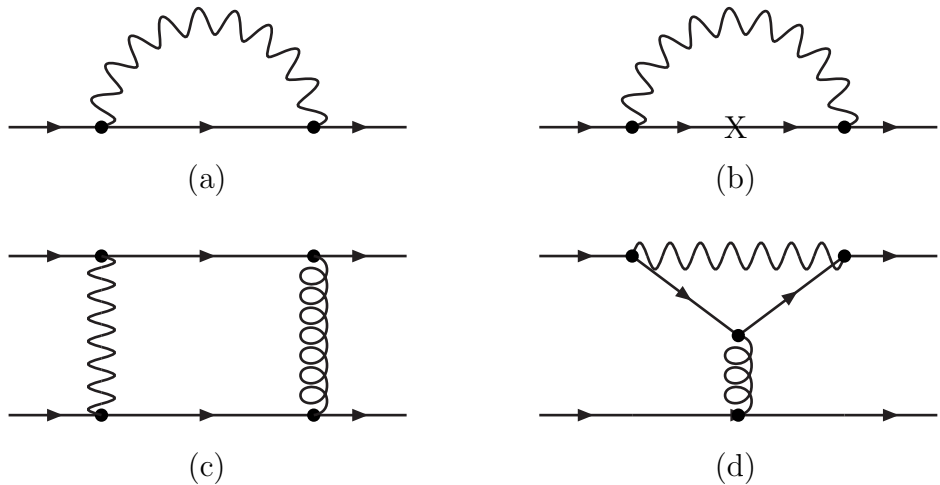

    In the leading order of $1/N_C$\footnote{The $1/N_c$ is a procedure where you 
      assume the number of coulors $N_C$ to be large so that you can expand in
      $1/N_C$.}, Fig.~\ref{htdtemc1}c and  Fig.~\ref{htdtemc1}d are just a product of 
    pure QCD currents and 
    their contributions have already been calculated in \cite{htdtemc2}. The 
    other two diagrams have to do with the renormalization of the scalar, 
    pseudo-scalar, vector and axial-vector currents, and the contribution depend 
    on the scale where the input current quark masses are renormalized in QED. This
    contribution has to some extent been calculated in \cite{htdtemc2}, but in 
    Ref.\cite{c2} by B. Moussallam it was pointed out that an extra contribution was 
    needed. This implies a different SD result for the $K_9$ to $K_{12}$ coefficients. 
    We will use the result from \cite{htdtemc2} as much as possible in this text
    and just add the ChPT corrections of $\mathcal{O}(p^4)$ which originates 
    from Fig.~\ref{htdtemc1}a and \ref{htdtemc1}b.
  
  \item[\bf{LD}]
    
    For this part we want to include all the orders in the QCD chiral Lagrangian 
    expansion, so that a proper and a smooth result will appear and all the 
    polynomial divergences will disappear. This is obviously not possible since 
    there are an infinite number of orders. The expansion up
    to $\mathcal{L}_6^{QCD}$ \cite{htdtemc1} is known, 
    but \emph{we will only consider $\mathcal{L}_2^{QCD}$ and $\mathcal{L}_4^{QCD}$}. 
    To this order the following relevant Feynman diagrams are generated:
  \begin{center}
    \begin{picture}
      (200,200)(0,40)
      \SetWidth{1.}
      
      \Line(10,50)(80,50)
 
      \GCirc(5,50){5}{1}
      \Line(10,45)(0,55) 
      \Line(0,45)(10,55) 
      
      \GCirc(85,50){5}{1}
      \Line(80,45)(90,55)
      \Line(90,45)(80,55)
      
      \PhotonArc(45,50)(15,0,180){2}{10}
      \GCirc(60,50){2}{0}
      \GCirc(30,50){2}{0}
      \Line(140,50)(220,50)
 
      \GCirc(145,50){5}{1}
      \Line(150,45)(140,55) 
      \Line(140,45)(150,55) 
      
      \GCirc(225,50){5}{1}
      \Line(220,45)(230,55)
      \Line(230,45)(220,55)
      
      \PhotonArc(185,60)(10,0,360){2}{10}
      \GCirc(185,52){2}{0}
      \Line(10,100)(80,100)
 
      \GCirc(5,100){5}{1}
      \Line(10,95)(0,105) 
      \Line(0,95)(10,105) 
      
      \GCirc(85,100){5}{1}
      \Line(80,95)(90,105)
      \Line(90,95)(80,105)
      
      \PhotonArc(5,114)(10,0,360){2}{8}
      \Line(150,100)(220,100)
 
      \GCirc(145,100){5}{1}
      \Line(150,95)(140,105) 
      \Line(140,95)(150,105) 
      
      \GCirc(225,100){5}{1}
      \Line(220,95)(230,105)
      \Line(230,95)(220,105)
      
      \PhotonArc(225,114)(10,0,360){2}{8}
      \Line(10,150)(80,150)
 
      \GCirc(5,150){5}{1}
      \Line(10,145)(0,155) 
      \Line(0,145)(10,155) 
      
      \GCirc(85,150){5}{1}
      \Line(80,145)(90,155)
      \Line(90,145)(80,155)
      
      \PhotonArc(25,150)(20,0,180){2}{10}
      \GCirc(45,150){2}{0}
      \Line(150,150)(220,150)
 
      \GCirc(145,150){5}{1}
      \Line(150,145)(140,155) 
      \Line(140,145)(150,155) 
      
      \GCirc(225,150){5}{1}
      \Line(220,145)(230,155)
      \Line(230,145)(220,155)
      
      \PhotonArc(205,150)(20,0,180){2}{10}
      \GCirc(185,150){2}{0}
      \GCirc(50,200){5}{1}
      \Line(45,195)(55,205) 
      \Line(55,195)(45,205)
      
      \GCirc(40,200){5}{1}
      \Line(35,195)(45,205)
      \Line(45,195)(35,205)
      
      \PhotonArc(45,215)(15,0,360){2}{10}
      \Line(150,200)(220,200)
 
      \GCirc(145,200){5}{1}
      \Line(150,195)(140,205) 
      \Line(140,195)(150,205) 
      
      \GCirc(225,200){5}{1}
      \Line(220,195)(230,205)
      \Line(230,195)(220,205)
      
      \PhotonArc(185,204)(40,0,180){2}{10}
    \end{picture}
  \end{center}
  The top left diagram might seem strange, but it just represents two external fields 
  coupled to a photon loop. Thus it consist out of one vertex and a photon propagator.

  Since we are only considering the leading and the next-to-leading order, problems 
  will appear. For the LD part these problems can be avoided by
  using low energies, i.e. a small $\Lambda$ which implies that terms like
  $L_i^2\frac{\Lambda^4}{F_0^4}$, $L_i^3\frac{\Lambda^6}{F_0^6}$, ... $\ll
  L_i\frac{\Lambda^2}{F_0^2}$. This means that to a good
  approximation for small momenta the higher order terms do not contribute.
  \end{itemize}
\end{description}

\begin{description}
\item[Right hand side, (RHS)]
  Here we have an effective Lagrangian in pure ChPT describing both QCD and QFD with
  internal photons. This
  means that ordinary field theory can be used. Since the tree level diagrams
  created by $\mathcal{L}_2$ are 
  \begin{center}
    \begin{picture}
      (240,20)(0,40)
      \SetWidth{1.}
      
      \Line(10,50)(70,50) 
      
      \GCirc(5,50){5}{1}
      \Line(0,45)(10,55) 
      \Line(10,45)(0,55) 
      
      \GCirc(75,50){5}{1}
      \Line(70,45)(80,55)
      \Line(80,45)(70,55)
      
      \GCirc(100,50){5}{1}
      \Line(95,45)(105,55) 
      \Line(105,45)(95,55) 
      
      \GCirc(110,50){5}{1}
      \Line(105,45)(115,55)
      \Line(115,45)(105,55)
      
      \Text(200,50)[]{$\mbox{of order } p^2 \mbox{ and } e^2$}
    \end{picture}
  \end{center}
  the vertices of $\mathcal{O}(p^2)$ can be put together to
  one loop diagrams of order $p^4, e^2p^2$ and $p^0e^4$. These diagrams generate
  divergences of the same order, which can be removed by adding tree level
  diagrams from 
  \begin{center}
    \begin{picture}
      (240,20)(0,40)
      \SetWidth{1.}
      
      \Line(10,50)(70,50) 
      
      \GCirc(5,50){5}{1}
      \Line(0,45)(10,55) 
      \Line(10,45)(0,55) 
      
      \GCirc(75,50){5}{1}
      \Line(70,45)(80,55)
      \Line(80,45)(70,55)
      
      \GCirc(100,50){5}{1}
      \Line(95,45)(105,55) 
      \Line(105,45)(95,55) 
      
      \GCirc(110,50){5}{1}
      \Line(105,45)(115,55)
      \Line(115,45)(105,55)
      
      \Text(220,50)[]{$\mathcal{L}_4 \mbox{ with }L_i \mbox{ of
          order } p^4$}
    \end{picture}
  \end{center}
  \begin{center}
    \begin{picture}
      (240,20)(0,40)
      \SetWidth{1.}
      
      \Line(10,50)(70,50) 
      
      \GCirc(5,50){5}{1}
      \Line(0,45)(10,55) 
      \Line(10,45)(0,55) 
      
      \GCirc(75,50){5}{1}
      \Line(70,45)(80,55)
      \Line(80,45)(70,55)
      
      \GCirc(100,50){5}{1}
      \Line(95,45)(105,55) 
      \Line(105,45)(95,55) 
      
      \GCirc(110,50){5}{1}
      \Line(105,45)(115,55)
      \Line(115,45)(105,55)
      
      \Text(220,50)[]{$\mathcal{L}_4$ $\mbox{with }K_i \mbox{ of
          order } e^2p^2$}
    \end{picture}
  \end{center}
  The $L_i$ coefficients will absorb the $p^4$ divergence from the loop
  diagram created
  by $\mathcal{L}_2$, and the $K_i$ coefficients will remove the $e^2p^2$
  divergence. \emph{Since we are only interested in the $K_i$ coefficients, the order
  that we are going to consider is $e^2p^2$.} By renormalizing with dimensional
  regularization and $\overline{MS}$ which is just a method to isolate the loop
  divergence and include some specific constant terms, all the infinities are 
  absorbed into the LECs (the $K_i$ and $L_i$ 
  coefficients in this case) and the Green
  function is well defined. When this renormalization is performed, a
  renormalization scale $\mu$ will appear just as in \cite{ntloel2}, this means that 
  in the end the LECs will depend on this scale. 
\end{description}

The resulting amplitudes from both sides are polynomials and logarithms of
momenta, mass and electric charge. The identifications are done particle by 
particle combined with different external currents. Since the loop integrals
are hard to calculate, simplifications, i.e putting $m_q=0$ or $p^2=0$ on both 
sides helps the calculations. The one-loop diagrams will generate infrared
(IR)\footnote{Just an other name for a something that go to infinity when
  the momentum goes to zero. The ultra violet (UV) divergences that show up in 
  the SD contribution, goes to infinity when the momentum goes to infinity.} 
divergences of the form $\log{\frac{\Lambda^2}{-p^2}}, 
\log{\frac{\Lambda^2}{m^2}}$ for the LHS and $\log{\frac{\mu^2}{-p^2}}, 
\log{\frac{\mu^2}{m^2}}$ for the RHS. As one can see this IR divergent part
will cancel on both sides. 

When both the SD and LD part on the LHS have been calculated and all the 
LECs of interest have been written in terms of $\Lambda$ the 
two parts are plotted against each other, and compared
to see at which region the result is approximately the same. In this way a 
reasonable region for the SD-LD separating scale $\Lambda$ can be determined, and 
it can then be removed.

\section{Result}

In this section the result for all the calculations are presented:
\newline

      
     
      
      
     
      
      
From the two-point function, using ChPT in the external field formalism, the LD 
part and the RHS can be calculated. After matching the amplitudes on the two sides 
the following relations are obtained:
\begin{eqnarray}
  \label{e1}
  C&=&\frac{3}{2}\frac{F^2_0\Lambda^2}{16\pi^2}+\frac{3L_{10}^r\Lambda^4}{16\pi^2}
\end{eqnarray}
Achieved through: $ \langle 0\mid T(\pi^+(x) \pi^+(y)) \mid 0\rangle$, using external
pseudoscalar currents in the chiral limit to the $\mathcal{O}(p^0)$.

\begin{eqnarray}
  \label{e2}
  K_8^r&=&\frac{3}{8\pi^2}\frac{\Lambda^2}{F^2_0}(4L_6^r-L_4^r)-\frac{8C}{F^4_0}
  (2L_6^r-L_4^r)=\frac{3}{8\pi^2}\frac{\Lambda^2L_4}{F^2_0}
\end{eqnarray}
Achieved through: $ \langle 0\mid T(\pi^+(x) \pi^+(y)) \mid 0\rangle$, using 
external pseudoscalar currents with $p^2=\hat{m}=0$ to the $\mathcal{O}(p^2)$.

\begin{eqnarray}
  \label{e3}
  K_9^r-K_8^r&=&\frac{1}{128\pi^2}\Big[(\xi-1)\log(\frac{\Lambda^2}{\mu^2})-
  \frac{3\Lambda^2}{F_0^2}(2L_4^r+L_5^r)\Big]
\end{eqnarray}
Achieved through: $ \langle 0\mid T(\pi^+(x) \pi^+(y)) \mid 0\rangle$, using 
external pseudoscalar currents with $p^2=m_s=0$ to the $\mathcal{O}(p^2)$.

\begin{eqnarray}
  \label{e4}
  K_9^r&=&\frac{1}{128\pi^2}\Big[(\xi-1)\log(\frac{\Lambda^2}{\mu^2})+
  \frac{3\Lambda^2}{F_0^2}(64L_6^r-18L_4^r-L_5^r)\Big]-\frac{8C}{F^4_0}
  (2L_6^r-L_4^r)
  \nonumber\\
  &=&\frac{1}{128\pi^2}\Big[(\xi-1)\log(\frac{\Lambda^2}{\mu^2})+
  \frac{3\Lambda^2}{F_0^2}(14L_4^r-L_5^r)\Big]
\end{eqnarray}
Achieved through: Eq.~\eqref{e2} and Eq.~\eqref{e3}.

\begin{eqnarray}
  \label{e5}
  4K_{11}^r+(K_4^r-2K_3^r)&=&\frac{1}{32\pi^2}
  \Big[(\xi+2)\log(\frac{\Lambda^2}{\mu^2})+
  \frac{1}{2}-12\frac{\Lambda^2}{F_0^2}L_9^r\Big]
\end{eqnarray}
Achieved through: $ \langle 0\mid T(\pi^+(x) \pi^+(y)) \mid 0\rangle$, using external
pseudoscalar currents in the chiral limit to the $\mathcal{O}(p^2)$.

\begin{eqnarray}
  \label{e6}
  2K_{12}^r-(K_4^r-2K_3^r)&=&\frac{1}{32\pi^2}\Big[-\log(\frac{\Lambda^2}{\mu^2})+
  \xi \log(\frac{\Lambda^2}{\mu^2})+6\frac{\Lambda^2}{F_0^2}L_9^r\Big]
\end{eqnarray}
Achieved through: $ \langle 0\mid T(\pi^+(x) \pi^+(y)) \mid 0\rangle$, using external
axial-vector currents in the chiral limit to the $\mathcal{O}(p^2)$.

\begin{eqnarray}
  \label{e7}
  2K_{11}^r+K_{12}^r&=&\frac{1}{64\pi^2}\Big[(1+2\xi)\log(\frac{\Lambda^2}{\mu^2})+
  \frac{1}{2}-6\frac{\Lambda^2}{F_0^2}L_9^r\Big]
\end{eqnarray}
Achieved through: Eq.~\eqref{e5} and Eq.~\eqref{e6}

\begin{eqnarray}
  \label{e8}
  -K_{13}^r+2K_{14}^r+4K_{12}^r-2(K_4^r-2K_3^r)&=&\frac{1}{32\pi^2}\Big[
  -2\log(\frac{\Lambda^2}{\mu^2})+\xi(\frac{-1}{4}+
  \frac{1}{2}\log(\frac{\Lambda^2}{\mu^2}))+
  \nonumber\\
  & &+\frac{12\Lambda^2}{F^2_0}L_9^r+
  \frac{\Lambda^2}{F_0^2}(L_{10}^r-2H_1^r)(12-3\xi)\Big]
\end{eqnarray}
Achieved through: $ \langle 0\mid T(\pi^+(x) \pi^+(y)) \mid 0\rangle$, using external
axial-vector currents in the chiral limit to the $\mathcal{O}(p^0)$.
 
\begin{eqnarray}
  \label{e9}
  \frac{2}{3}(K_5^r+K_6^r)&=&-(K_4^r-2K_3^r)=-2(K_1^r+K_2^r)
  \\
  \label{e10}
  K_7^r+K_8^r&=&0
  \\
  \label{e11}
  K_9^r+K_{10}^r&=&0
\end{eqnarray}

In the calculation of the coefficients from the charged currents in 
Eq.~\eqref{e2} to Eq.~\eqref{e8} we have used the result from the neutral currents in 
Eq.~\eqref{e9} to Eq.~\eqref{e11}. The observant reader has probably already noticed 
that some of the coefficients above are linear combinations of others. All together 
there exist 10 real combinations, the ones that are linear combinations are 
Eq.~\eqref{e4}, Eq.~\eqref{e7} and  Eq.~\eqref{e9}. The reason why we display the 
linear combinations is that only certain combinations of the EM LECs have been 
calculated before, and to achieve those we need to rearrange our coefficients. 

The processes that we used to calculate the coefficients above are not the only ones
investigated, e.g. for the charge current we also investigated $K^+$, but the 
combinations that came out of those calculations gave the same result as for the 
$\pi^+$. For the neutral current we investigated contributions from $\eta$, $\pi^0$, 
$\bar{K}^0$ and $K^0$, which all contributed to the combinations in Eq.~\eqref{e9} to 
Eq.~\eqref{e11}.
\newline

For the SD part which we calculated with perturbative QCD we got after comparing to 
the RHS:

\begin{eqnarray}
  C&=&\frac{3\alpha_s}{8\pi}\frac{F^4_0B_0^2}{\Lambda^2}
  \\
  K_1^r&=&\frac{3\alpha_s}{32\pi}\frac{F_0^2}{\Lambda^2}-
  \frac{2}{3\pi}\frac{L_4^rB_0^2}{\Lambda^2}
  \\
  K_2^r&=&0
  \\
  K_3^r&=&\frac{-3\alpha_s}{32\pi}\frac{F_0^2}{\Lambda^2}
  \\
  K_4^r&=&0
  \\ 
  K_5^r&=&-\frac{\alpha_s}{6\pi}\frac{F_0^2}{\Lambda^2}-
  \frac{\alpha_s}{3\pi}\frac{L_5^rB_0^2}{\Lambda^2}
  \\ 
  K_6^r&=&\frac{\alpha_s}{2\pi}\frac{L_5^rB_0^2}{\Lambda^2}
  \\
  K_7^r&=&\frac{-4\alpha_s}{3\pi}\frac{L_6^rB_0^2}{\Lambda^2}
  \\
  K_8^r&=&\frac{6\alpha_s}{\pi}\frac{L_6^rB_0^2}{\Lambda^2}
\end{eqnarray}
\begin{eqnarray}
  K_9^r&=&\frac{-\alpha_s}{6\pi}\frac{(2L_8^r+H_2^r)B_0^2}{\Lambda^2}-
  \frac{(1-\xi)}{128\pi^2}\log{\frac{\mu^2_0}{\Lambda^2}}
  \\ 
  K_{10}^r&=&\frac{3\alpha_s}{4\pi}\frac{(2L_8^r+H_2^r)B_0^2}{\Lambda^2}+
  \frac{(4-\xi)}{128\pi^2}\log{\frac{\mu^2_0}{\Lambda^2}}
  \\
  K_{11}^r&=&\frac{3\alpha_s}{4\pi}\frac{(2L_8^r-H_2^r)B_0^2}{\Lambda^2}-
  \frac{(4-\xi)}{128\pi^2}\log{\frac{\mu^2_0}{\Lambda^2}}
  \\
  K_{12}^r&=&\frac{(\xi-1)}{64\pi^2}\log{\frac{\mu^2_0}{\Lambda^2}}
\end{eqnarray}
where the log contribution comes from Fig.~\ref{htdtemc1}a and Fig.~\ref{htdtemc1}b. 

To be able to plot the SD part against the LD part, all the relevant coupling 
constants have to be known and the renormalization scales have to be chosen. 
All this is displayed in Tab.~\ref{cc1}.
\begin{table}[h]
  \label{cc1}
  \begin{center}
    \begin{tabular}{|c|c|c|c|}
      \hline
      Coefficient & fit 10 & $\mathcal{O}(p^4)$ & article\\
      \hline
      $10^3L_1^r$&0.43$\pm$ 0.12&0.38&\cite{a2}\\
      \hline
      $10^3L_2^r$&0.73$\pm$ 0.12&1.59&\cite{a2}\\
      \hline
      $10^3L_3^r$&-2.35$\pm$ 0.12&-2.91&\cite{a2}\\
      \hline
      $10^3L_4^r$&0&0&\cite{a2}\\
      \hline
      $10^3L_5^r$&0.97$\pm$ 0.11&1.46&\cite{a2}\\
      \hline
      $10^3L_6^r$&0&0&\cite{a2}\\
      \hline
      $10^3L_7^r$&-0.31$\pm$ 0.14&-0.49&\cite{a2}\\
      \hline
      $10^3L_8^r$&0.60$\pm$ 0.18&1.00&\cite{a2}\\
      \hline
      $10^3L_9^r$&5.93$\pm$ 0.43&6.0&\cite{a1}\\
      \hline
      $10^3L_{10}^r$&&-5.5&\cite{ntloel1}\\
      \hline
      $10^3(2L_8^r+H_2^r)$&&2.9$\pm$ 1.0&\cite{a1}\\
      \hline
      $10^3H_1^r$&&$-4.1$&\\
      \hline
      $F_0[MeV]$&87.7&81.1&\cite{a2}\\
      \hline
    \end{tabular} 
    \begin{tabular}{|c|c|}
      \hline
      $B_0[GeV]$&1.63\\
      \hline
      $\alpha_s$&0.3\\
      \hline
      $\mu[GeV]$ (ChPT scale)&0.77\\
      \hline
      $\mu_0[GeV]$ (QCD scale)&1.0\\
      \hline
    \end{tabular}
  \end{center}
\caption{Displays the coupling constants that are needed to calculate the $K_i$s. 
The finite renormalized parts are at the scale $M_\rho=0.77$ $GeV$. The values of 
the parameters to the left are all determined from articles. As one can see there 
are missing three fit 10 values. For these we use $\mathcal{O}(p^4)$ in both cases. 
The 
parameters to the right have values that we have chosen ourselves. $B_0$ are on 
the QCD scale $\sim$ 1 $GeV$ and $\alpha_s$ is assumed to be constant. The ChPT 
parameter $\mu$ that appeared when we performed the dimensional regularization 
is chosen to be $\mu=M_\rho=0.77$ $GeV$.}
\end{table}
As one might have noticed there are large errors in most of the parameters, and 
that will of course affect our result in the end. This is the reason why we have 
chosen two different fitting methods, fit 10\footnote{Include terms up to 
  $\mathcal{O}(p^6)$, and is based on that $L_4^r$ and $L_6^r$ is equivalent to zero.}
and $\mathcal{O}(p^4)$\footnote{Include terms up to   $\mathcal{O}(p^4)$, and is 
based on that $L_4^r$ and $L_6^r$ is equivalent to zero}, respectively, 
i.e. the change in the result in the end can be seen. This will of course affect some 
$K_i^r$ constants more than others. In our calculations we also fix the gauge of 
the photon and use Feynman gauge, which means that the gauge fixing 
term $\xi$ is put equal to zero. 

\begin{figure}[h!]
  \begin{center}
    \includegraphics[angle=270, scale=0.43]{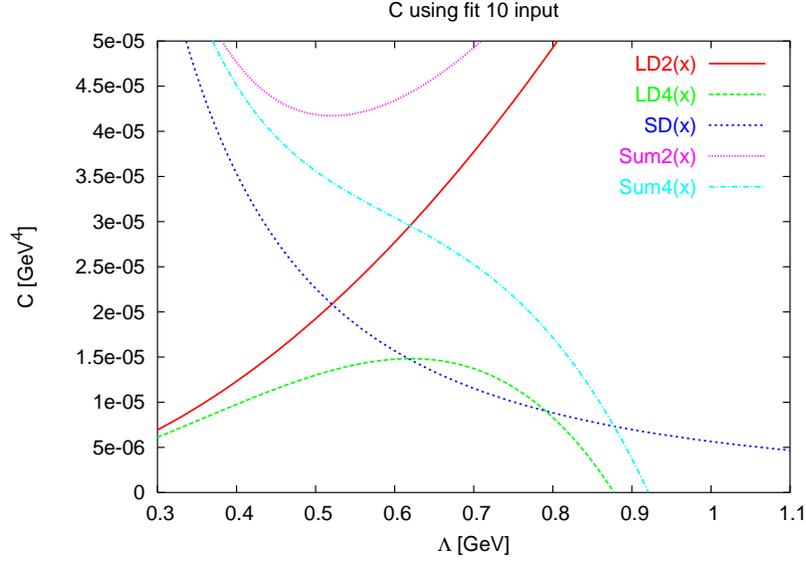}
  \end{center}
  \caption{The long distance (LD), short distance (SD) and the sum as 
    a function of the matching variable $\Lambda$ for C. The number after the 
    functions represents which $\mathcal{O}(p^n)$ that are considered. The fit 10 
    input parameters are used.}
  \label{cfit}
\end{figure}

\begin{figure}[h!]
  \begin{center}
    \includegraphics[angle=270, scale=0.43]{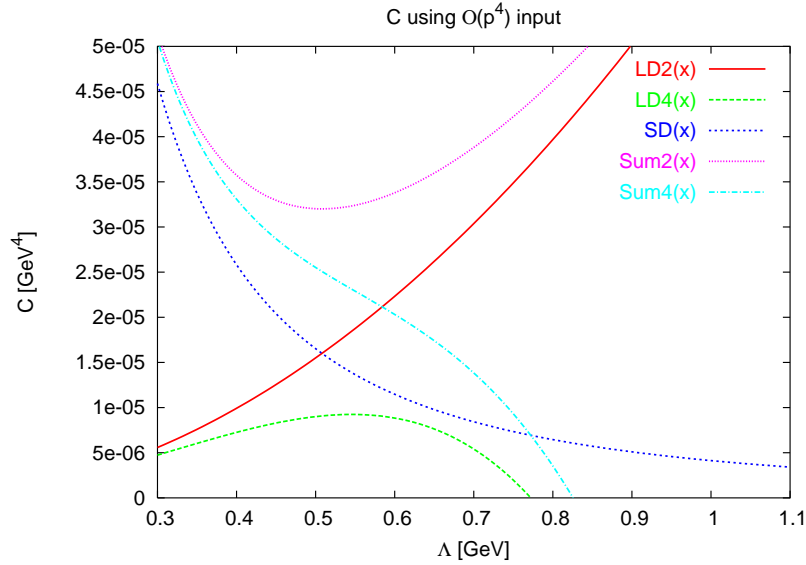}
  \end{center}
  \caption{The long distance (LD), short distance (SD) and the sum as 
    a function of the matching variable $\Lambda$ for C. The number represents 
    which $\mathcal{O}(p^n)$ that are considered. The $\mathcal{O}(p^4)$ input 
    parameters are used.}
  \label{cop4}
\end{figure}

\begin{figure}[h!]
  \begin{center}
    \includegraphics[angle=270, scale=0.43]{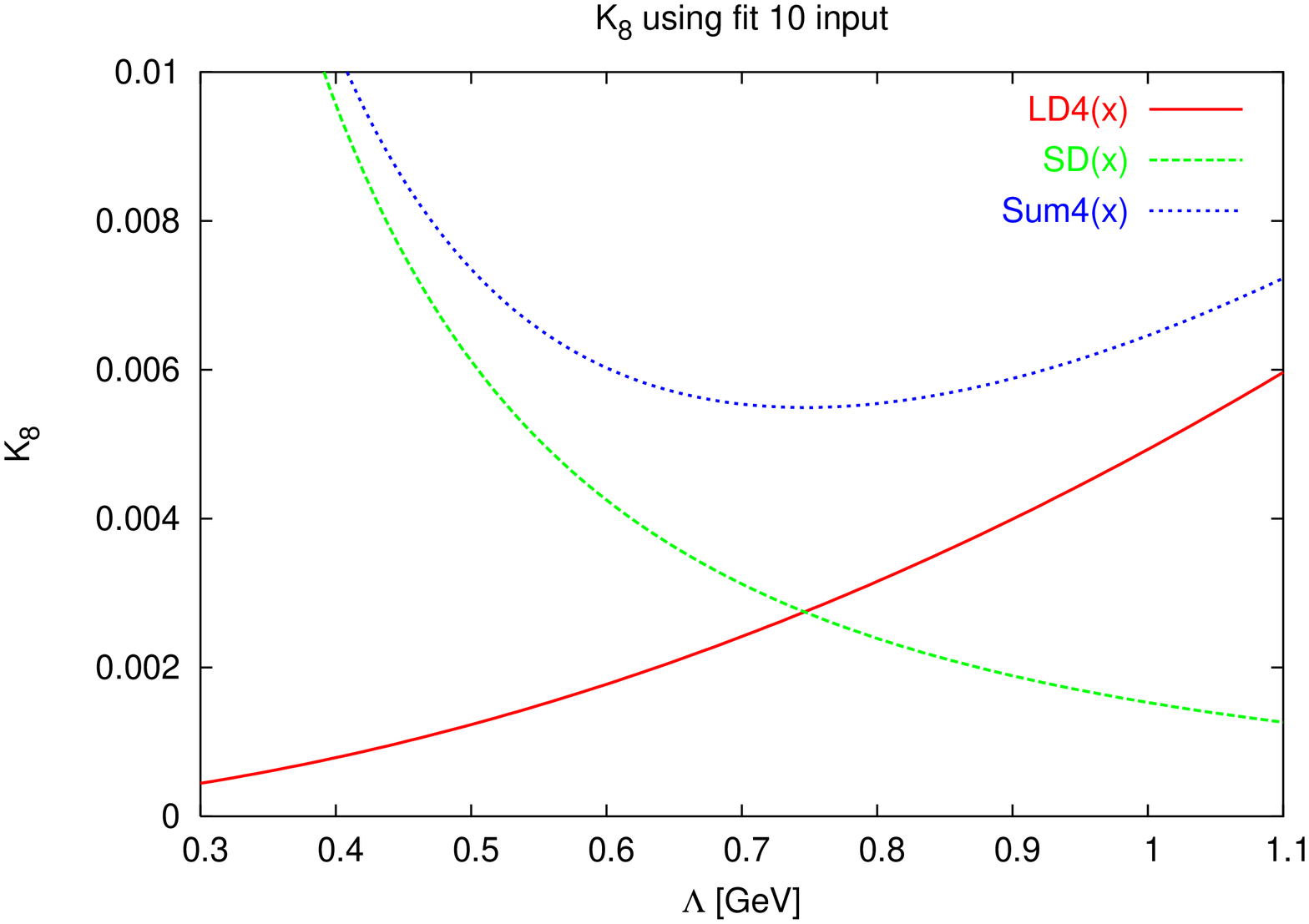}
  \end{center}
  \caption{The long distance (LD), short distance (SD) and the sum as 
    a function of the matching variable $\Lambda$ for $K_8^r$. The number 
    represents which $\mathcal{O}(p^n)$ that are considered. The fit 10 input 
    parameters are used.}
  \label{k8fit}
\end{figure}

\begin{figure}[h!]
  \begin{center}
    \includegraphics[angle=270, scale=0.43]{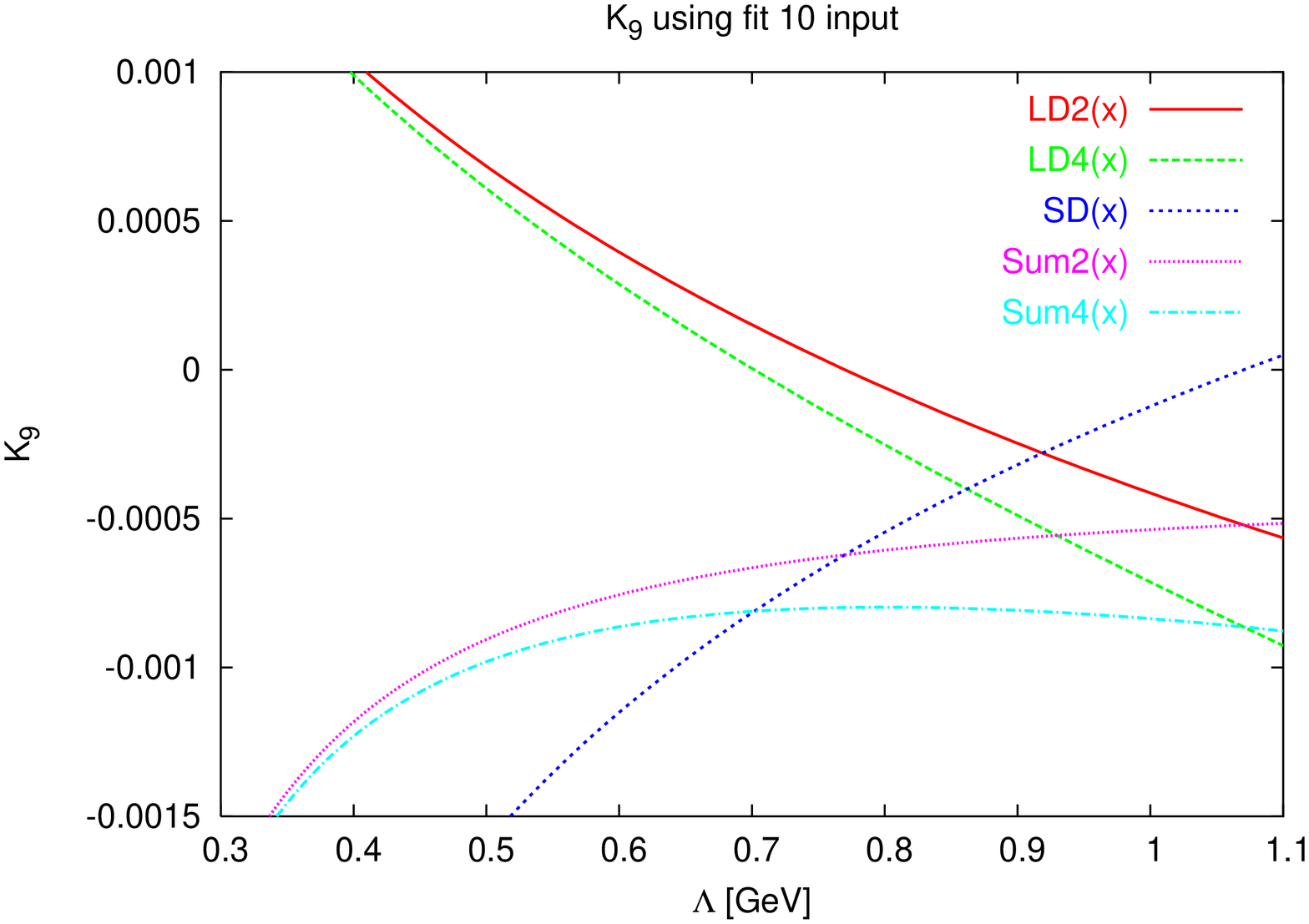}
  \end{center}
  \caption{The long distance (LD), short distance (SD) and the sum as 
    a function of the matching variable $\Lambda$ for $K_9^r$. The number 
    represents which $\mathcal{O}(p^n)$ that are considered. The fit 10 input 
    parameters are used.}
    \label{k9fit}
\end{figure}

\begin{figure}[h!]
  \begin{center}
    \includegraphics[angle=270, scale=0.42]{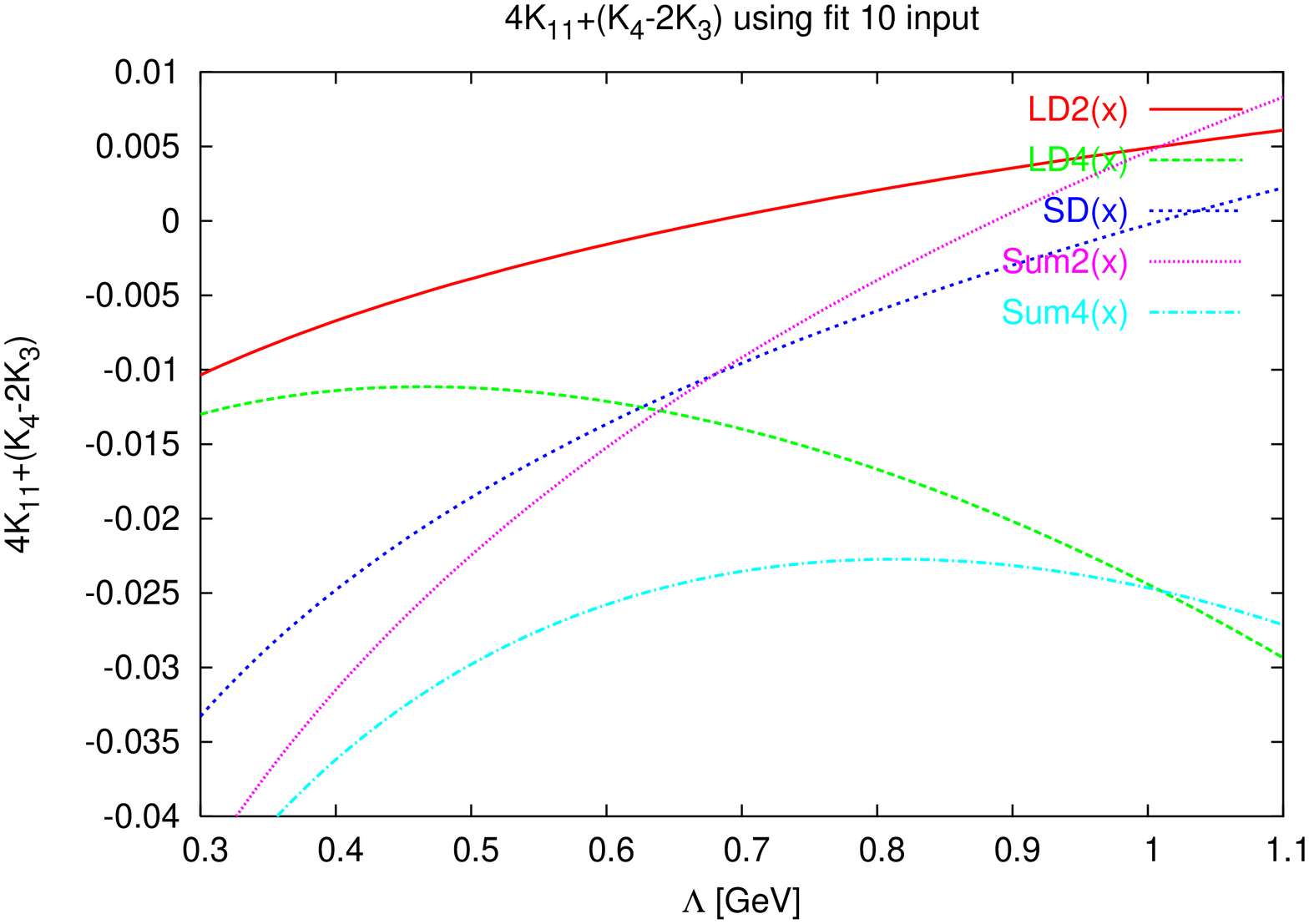}
  \end{center}
  \caption{The long distance (LD), short distance (SD) and the sum as 
    a function of the matching variable $\Lambda$ for $4K_{11}^r+(K_4^r-2K_3^r)$. 
    The number after the 
    functions represents which $\mathcal{O}(p^n)$ that are considered. The fit 10 
    input parameters are used.}
  \label{k11fit}
\end{figure}

\begin{figure}[h!]
  \begin{center}
    \includegraphics[angle=270, scale=0.42]{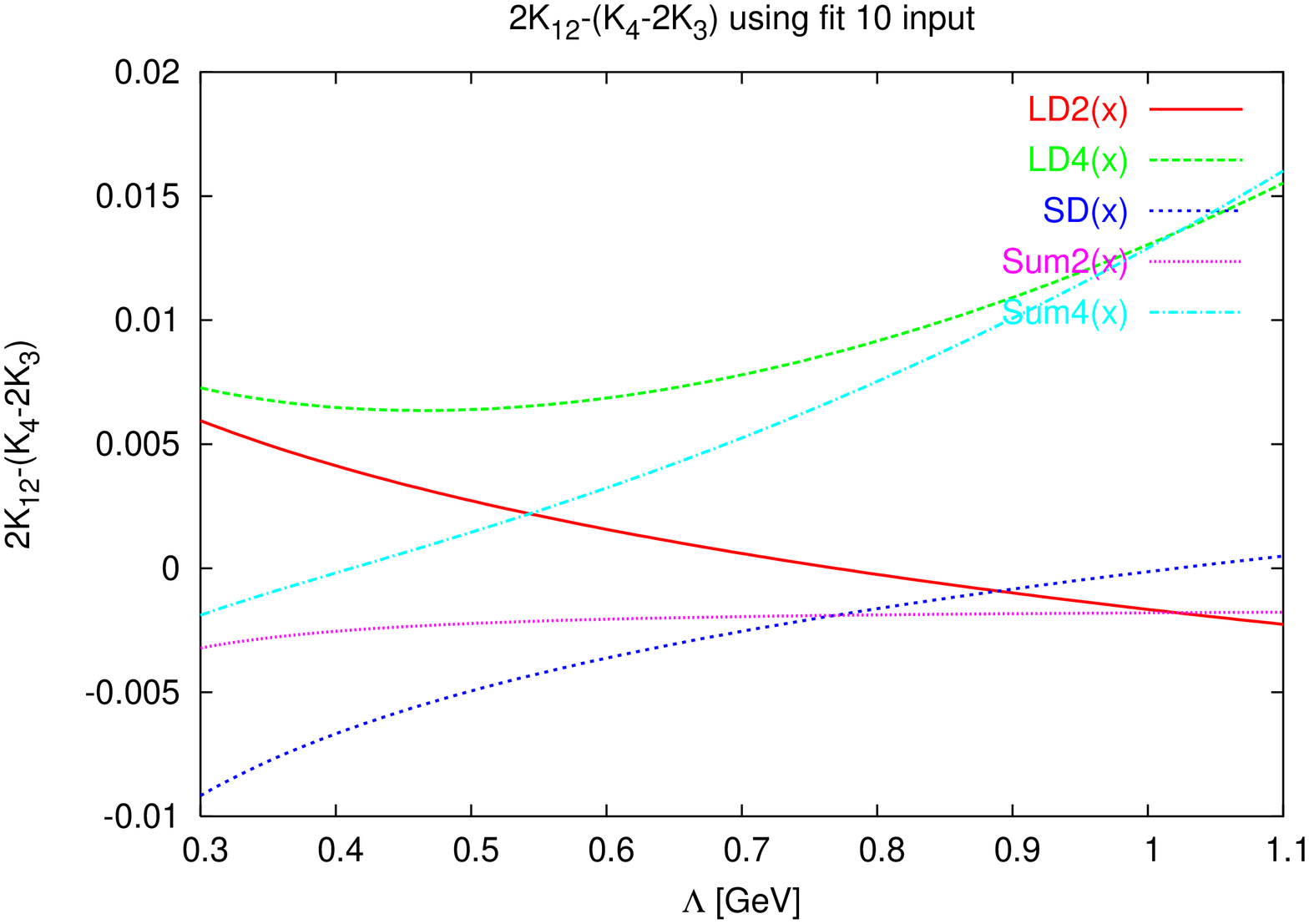}
  \end{center}
  \caption{The long distance (LD), short distance (SD) and the sum as 
    a function of the matching variable $\Lambda$ for $2K_{12}^r-(K_4^r-2K_3^r)$. 
    The number after the 
    functions represents which $\mathcal{O}(p^n)$ that are considered. The fit 10 
    input parameters are used.}
  \label{k12fit}
\end{figure}

\begin{figure}[h!]
  \begin{center}
    \includegraphics[angle=270, scale=0.43]{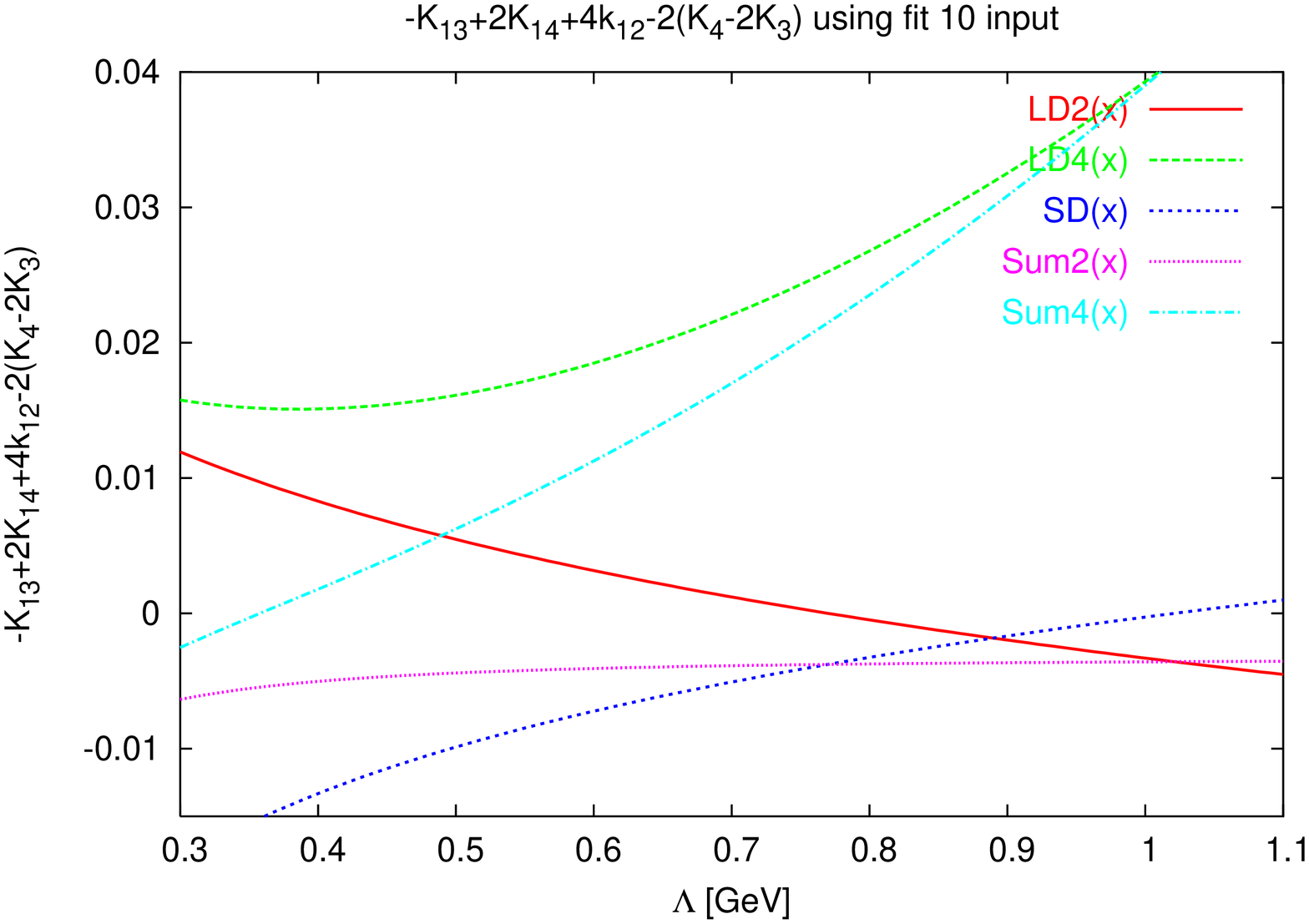}
  \end{center}
  \caption{The long distance (LD), short distance (SD) and the sum as 
    a function of the matching variable $\Lambda$ for $K_{13}^r+2K_{14}-
    2(K_4^r-2K_3^r)$. The number after the 
    functions represents which $\mathcal{O}(p^n)$ that are considered. The fit 10 
    input parameters are used.}
  \label{k13fit}
\end{figure}

\begin{figure}[h!]
  \begin{center}
    \includegraphics[angle=270, scale=0.43]{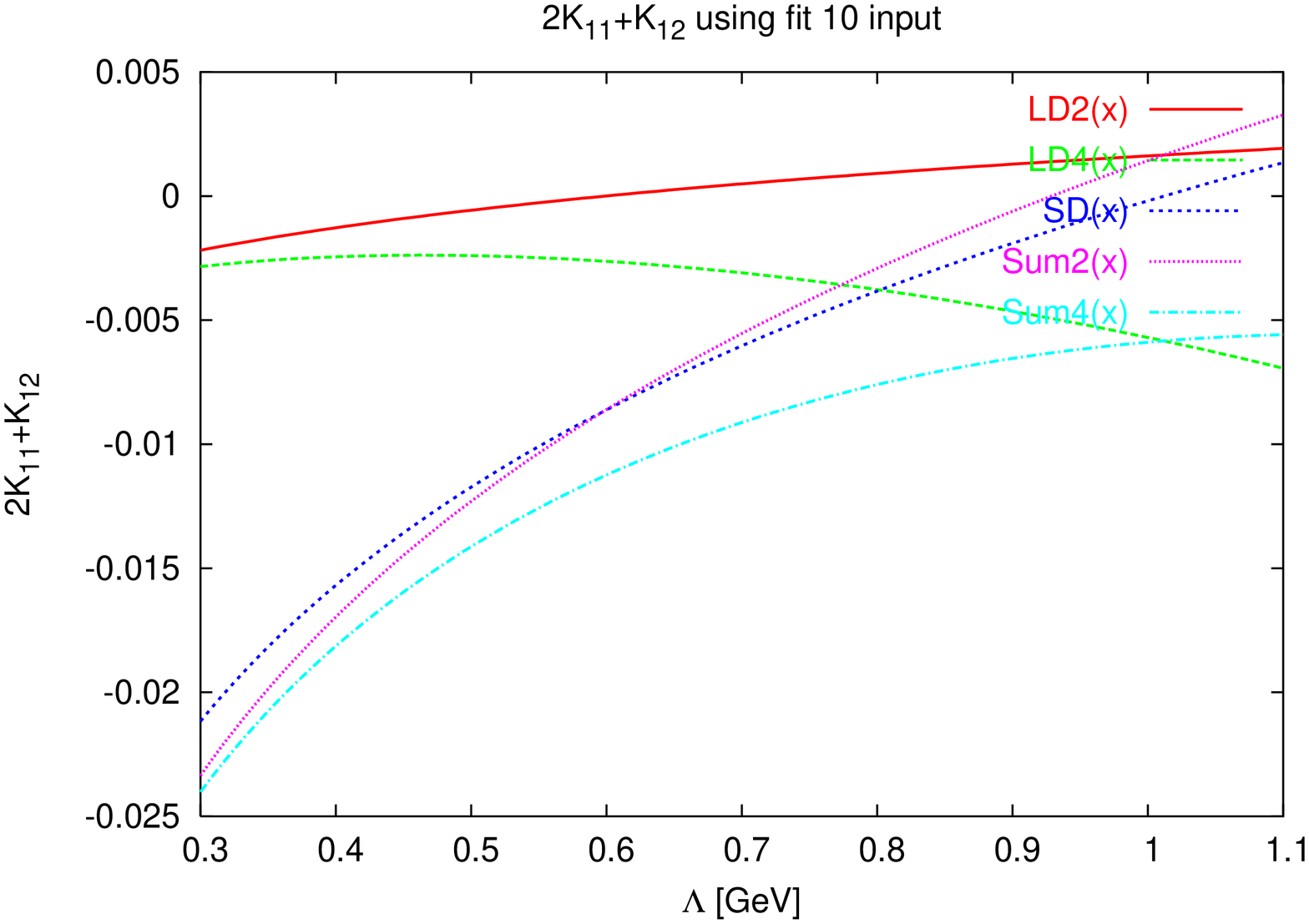}
  \end{center}
  \caption{The long distance (LD), short distance (SD) and the sum as 
    a function of the matching variable $\Lambda$ for $2K_{11}^r+K_{12}$. The number 
    after the 
    functions represents which $\mathcal{O}(p^n)$ that are considered. The fit 10 
    input parameters are used.}
  \label{k112fit}
\end{figure}

All that remains now is to compare the LD with the SD, component by component, and 
that is done in Figs.~\ref{cfit} to Fig.~\ref{k112fit}. On the y-axis we have the 
value of the LEC and on the x-axis we have the cut-off $\Lambda$. The scale on the 
x-axis range from 0.3 $GeV$ to 1.1 $GeV$, and the reason for this is that a
$\Lambda$ with a too high energy ($\sim$ 1 $GeV$) would disturb the 
perturbative treatment (can be seen from the figures where the higher order deviates 
faster for bigger $\Lambda$) and since our ChPT scale is around 1 $GeV$, it is not 
proper to exceed that energy. The reason why at low energies it does not work is that 
the perturbative expansion is not valid below $\sim$ 0.3 $GeV$.

The matching is done by finding the best region for 
LD4 and SD. This is done by looking at the region 
where the Sum4 function has the derivative closest 
to zero, but one also has to consider how the LD and SD parts behave. From 
Fig.~\ref{cfit} and Fig.~\ref{cop4} describing the coefficient C one can see how the 
next order (LD4) affects the result. If only the lowest order LD term 
were considered, the matching region would be easy to find. By adding the 
next order one can see that the matching region is not so good anymore, which means 
that the result would have a larger error. The difference between LD2 and LD4 is 
important, since in this text we are also interested in comparing the behavior of the 
two functions. As one can see there is no big 
difference between the fit 10 input and the $\mathcal{O}(p^4)$ input, that is why we 
only include the fit 10 graphs for the remaining quantities.

When $K_8^r$ is calculated as in Fig.~\ref{k8fit} the $L_4^r$ and the $L_6^r$ 
were put equal to $10^{-3}$ to avoid a non-zero result. In the figure 
we see that the sum of the LD and SD behaves nicely, which is good.

For $K_9$ in Fig.~\ref{k9fit} both the sums behave nicely. The best matching region 
is a little too high, but it is still good enough. When we compare 
the two sums, we see that Sum4 is a little better.

For the $4K_{11}^r+(K_4^r-2K_3^r)$ combination in Fig.~\ref{k11fit} the Sum4 function 
behaves in an appealing way. The behavior of Sum2 is also rather good, the only thing 
to complain about is that the zero derivative area is a bit too low for Sum2 and a bit 
too high for Sum4. There is also a large difference between them.

Next we consider Fig.~\ref{k12fit}, Fig.~\ref{k13fit} and Fig.~\ref{k112fit} 
describing $2K_{12}^r-(K_4^r-2K_3^r)$ , $-K_{13}^r+2K_{14}^r+4K_{12}-2(K_4^r-2K_3^r)$ 
and $2K_{11}^r+K_{12}^r$, respectively. These graphs do not have a Sum4 that behaves 
as 
we want in the region of interest, but we still pick a $\Lambda$ in our region, which 
means that the error in these parameters are big. For the $2K_{12}^r-(K_4^r-2K_3^r)$ 
and 
$-K_{13}^r+2K_{14}^r+4K_{12}-2(K_4^r-2K_3^r)$ combinations one can also see that the 
Sum2 behaves much better.
\clearpage
\clearpage
\clearpage
\clearpage

The result from all the figures are summerized in the Tab.~\ref{t2} below:\\
\begin{table}[h!]
  \begin{center}
    \caption{The results for fit 10, ``This work 1'' is the value calculated in this 
      text using fit 10, for ``This work 2'' we are using $\mathcal{O}(p^4)$. 
      The $\Lambda$ 
      is value of the cut-off where the SD part best matches the LD part.}
    \label{t2}
    \begin{tabular}{|c|c|c|c|c|}
      \hline
      Coefficient & This work 1 & $\Lambda$, fit 10&This work 2 &$\Lambda$, 
      $\mathcal{O}(p^4)$\\
      &&[$GeV$]&&[$GeV$]\\
      \hline
      C&$3.0\times 10^{-5}$ &$0.62$&$2.25\times 10^{-5}$&$0.58$\\
      &[$GeV^4$]&&[$GeV^4$]&\\
      \hline
      $K_8^r$ (used $L_4^r=L_6^r=10^{-3}$)&$5.8\times 10^{-3}$&$0.72$&
      $6\times 10^{-3}$&$0.71$\\
      \hline
      $K_9^r$ &$-8\times 10^{-4}$&$0.8$&$-9\times 10^{-4}$&$0.73$\\
      \hline
      $4K_{11}^r+(K_4^r-2K_3^r)$ &$-2.3\times 10^{-2}$&$0.83$&$-2.3\times 10^{-2}$
      &$0.71$\\
      \hline
      $2K_{12}^r-(K_4^r-2K_3^r)$ &$4\times 10^{-3}$&$0.65$&$5\times 10^{-3}$&$0.65$\\
      \hline
      $-K_{13}^r+2K_{14}^r+4K_{12}-2(K_4^r-2K_3^r)$ &$1.1\times 10^{-2}$&$0.6$&
      $1.4\times 10^{-2}$&$0.6$\\
      \hline
      $2K_{11}^r+K_{12}^r$&$-8\times 10^{-3}$&$0.8$&$-8\times 10^{-3}$&$0.8$\\
      \hline
    \end{tabular}
  \end{center}
\end{table}
\begin{table}[h!]
  \begin{center}
    \caption{Displays the relevant calculated coefficients for each article. For 
      \cite{htdtemc2} $F_0=89MeV$ was used, for \cite{c2} the improved Z value for 
      $K_9$ and $K_{10}$ was used. In all the calculation of 
      the values we have used $\mu_0=1$ $GeV$, $M_V=\mu=M_\rho=0.77$ $GeV$, 
      $F_\pi=F_0=92.4$ $MeV$, $M_A=M_{a_1}=1.23$ $GeV$ and leading $N_C$ order.}
    \label{t1}
    \begin{tabular}{|c|c|c|c|}
      \hline
      Coefficient & \cite{htdtemc2} & \cite{c2} & \cite{c3}\\
      \hline
      C  [$GeV^4$]&$4.2\times 10^{-5}$&$2.36\times 10^{-5}$&\\
      \hline
      $K_8^r$ &$0.8\times 10^{-3}$&0&\\
      \hline
      $K_9^r$ &$-1.3\times 10^{-3}$&$-3.62\times 10^{-3}$&\\
      \hline
      $4K_{11}^r+(K_4^r-2K_3^r)$ &$-5.0\times 10^{-3}$&&\\
      \hline
      $2K_{12}^r-(K_4^r-2K_3^r)$ &&&\\
      \hline
      $-K_{13}^r+2K_{14}^r+4K_{12}-2(K_4^r-2K_3^r)$&&&\\
      \hline
      $2K_{11}^r+K_{12}^r$&&$-5\times 10^{-3}$&$-1.94\times 10^{-3}$\\
      \hline
    \end{tabular}
  \end{center}
\end{table}

The relevance of some of the results in Tab.~\ref{t2} is now discussed. 

When we compare the values for C in Tab.~\ref{t2} with the values in the articles in 
Tab.~\ref{t1}, one can see that they are rather simular, which is good. 

The value for $K_8^r$ will not be discussed because of the arbitrary chosen $L_i^r$ 
parameters. 

Instead we turn to $K_9^r$ which have a good matching region, and as we see these 
values are rather close to the one derived in \cite{htdtemc2}. 

The values for $4K_{11}^r+(K_4^r-2K_3^r)$ are only calculated in \cite{htdtemc2}, 
and they differ a lot compared to our values. Why is hard to say, but there are a 
lot of approximations that have been done in both cases.

For $2K_{12}^r-(K_4^r-2K_3^r)$ and $-K_{13}^r+2K_{14}^r+4K_{12}-2(K_4^r-2K_3^r)$ no 
published values are accessible. So their values will not be commented further.

The last combination obtained is $2K_{11}^r+K_{12}^r$, which has values that are 
close to the value 
calculated in  \cite{c2}. This means that our values are reasonable\footnote{To say 
that our value is reasonable, does not mean that it is the right value, since we do 
not really know which value that is the right one.}, despite the large uncertainty.

In Tab.~\ref{t2} one can see that the fit 10 and $\mathcal{O}(p^4)$ input do not differ
 much, which means that the largest error in the $K_i^r$ coefficients are not due to 
the uncertainties in the $L_i^r$s, but instead due to the matching between the LD and 
SD parts.

\section{Conclusion}

In this thesis we have determined the renormalized and finite coefficients of the
generating functional of ChPT up to $\mathcal{O}(p^4)$ including virtual 
photons to one-loop. The finite part incorporates 29 low energy coupling constants,
and of those, 10 combinations of some the 14 $K_i^r$s have in the $\overline{MS}$ 
scheme been written in 
terms of the experimentally known constants $L_i^r$, the cut-off $\Lambda$ and the
dimensional regularization parameter $\mu$. The $\Lambda$ dependences have been 
removed through the matching of the LD part to the SD part, which had a good 
match for some of the coefficients and a not so good for others. But over all the
$\mathcal{O}(p^4)$ contribution made the result a little bit better. But the 
improvement
 is still not that big as we hoped it would be, which implies that the result 
of the naive leading order estimate in \cite{c1} is often good enough for a first
estimate in the determination of the $K_i$ constants. One could argue that a 
contribution of $\mathcal{O}(p^6)$ would improve this result, but it seems 
unlikely since higher orders tend to deviate more for $\Lambda$s in the upper region. 
\newline

The determination of the effective low energy chiral Lagrangian is of great 
importance since it is used to determine the free quark mass difference $m_u-m_d$, 
which is not very well known. The procedure used, is based on that the physical 
$m_{K^+}-m_{K^0}$ mass difference has a pure QCD contribution due to the 
difference in quark masses and an electromagnetic contribution. Since the mass of 
the mesons can be determined experimentally, and 
quite accurate, then 
by calculating the EM contribution, the quark mass difference can be determined. 
This contribution can be calculated through different ChPT techniques, some better 
than others, often depending on region of interest. The procedure used by 
Bijnens-Prades in \cite{htdtemc2} includes all orders of $p^2$ in ChPT for the LD 
part, and this is done by using the ENJL model. Ananthanarayan-Moussallam in 
\cite{c2} and \cite{c3} also used a 
method where all orders of $p^2$ in ChPT were considered. The difference is that 
Moussallam used resonance saturation instead, but we will not explore the two methods 
more since it is beyond of the scope of this thesis.

One of the problems calculating the next-to-leading order electromagnetic coefficients 
is that the QCD coefficients are needed, the $L_i^r$s in this case. Most of these 
terms are not so well determined. In the near future, hopefully, these constants 
can be determined more accurately.
\newline

Using the ChPT expansion on two-point functions as we did for the LD part, was not 
enough to determine all the LECs, as we were hoping. We even did the calculations
for the four-point function for the $K^0$ ($\overline K^0$), but no new relation 
appeared. In the future one should investigate the four-point functions further and 
see if new $K_i^r$ relations drop out. 

 
\clearpage
\appendix

\section{An easy example}

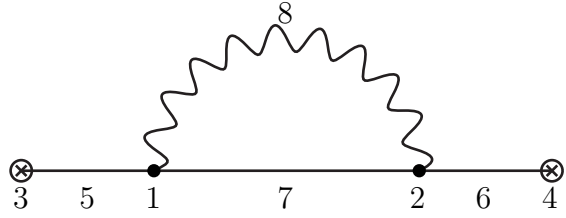
\begin{figure}
  \begin{center}
    \begin{picture}
      (215,140)(-5,0)
      \SetWidth{1.}
      
      \Line(0,50)(200,50)
      \PhotonArc(100,50)(50,0, 180){5}{10}
      \Text(0,40)[]{3}
      \Text(25,40)[]{5}
      \Text(50,40)[]{1}
      \Text(100,40)[]{7}
      \Text(150,40)[]{2}
      \Text(175,40)[]{6}
      \Text(200,40)[]{4}
      \Text(100,110)[]{8}
      
      \CArc(0,50)(4,0,360)
      \Line(2,52)(-2,48)
      \Line(2,48)(-2,52)
      \CArc(200,50)(4,0,360)
      \Line(198,48)(202,52)
      \Line(202,48)(198,52)
      
      \GCirc(50,50){2}{0}
      \GCirc(150,50){2}{0}
    \end{picture}
  \end{center}
  \caption{One of the basic Feynman diagrams for a two-point Green function.}
  \label{fig:ex1}
\end{figure}

Start by considering the leading-order effective QCD Lagrangian in
Eq.~\eqref{eqqcdale1}. The $U$ field in Eq.~\eqref{eqloel2}and Eq.~\eqref{eqloel3} 
is expanded in the following way:
\begin{align}
  \label{eqaee1}
  U= \exp(\frac{i\Phi}{F_0}) \simeq1+i\frac{\Phi}{F_0}-\frac{\Phi^2}{2F_0^2}+\cdots
  \nonumber\\
  U^\dagger= \exp(\frac{i\Phi}{F_0}) \simeq1-i\frac{\Phi}{2F_0}-
  \frac{\Phi^2}{F_0^2}+\cdots
\end{align}
where $\Phi^\dagger=\Phi$ has been used. Taking the partial derivative of U in 
Eq.~\eqref{eqaee1} yields
\begin{align}
  \label{eqaee2}
  &\partial_\mu U=\frac{i}{F_0} \partial_\mu \Phi-\frac{1}{2F_0}(\partial_\mu
  \Phi \Phi+\Phi \partial_\mu \Phi)
  \nonumber\\
  &\partial_\mu U^\dagger=-\frac{i}{F_0} \partial_\mu \Phi-\frac{1}{2F_0}(
  \partial_\mu \Phi \Phi+\Phi \partial_\mu \Phi).
\end{align}
Putting Eq.~\eqref{eqaee2} into the covariant derivative in Eq.~\eqref{eqloel4} gives
\begin{align}
  &D_\mu U=\partial_\mu U-ieQA_\mu U+iUeQA_\mu=
  \nonumber\\
  &\frac{i}{F_0} \partial_\mu \Phi-
  \frac{1}{2F_0^2}(\partial_\mu \Phi \Phi+\Phi \partial_\mu \Phi)-
  ie[QA_\mu,\frac{i}{F_0}\Phi-\frac{\Phi^2}{F_0^2}]+\cdots
  \nonumber\\
  &D_\mu^\dagger U=-\frac{i}{F_0} \partial_\mu \Phi-
  \frac{1}{2F_0^2}(\partial_\mu \Phi \Phi+\Phi \partial_\mu \Phi)-
  ie[QA_\mu,\frac{i}{F_0}\Phi-\frac{\Phi^2}{F_0^2}]+\cdots
\end{align}
Now an expression for the first term in the Lagrangian in Eq.~\eqref{eqqcdale1} has
been derived and the Feynman diagram in Fig.~\ref{fig:ex1} is ready to be
evaluated.

\begin{enumerate}
\item
  Here a three vertex is investigated, where the only relevant terms from
  Eq.~\eqref{eqqcdale1} are the ones with one photon field $A_\mu$, one $\Phi$ term and
  one $\partial_\mu \Phi$ term. So 
  \begin{align}
    &tr(D_\mu U D^\mu U)=\cdots=\frac{-2eiA^\mu }{F_0^2}tr(\pmu \Phi[Q,\Phi])+\cdots
    =\cdots=
    \nonumber\\
    &\frac{-2eiA^\mu}{F_0^2}(-\pmu \pi^+ \pi^--\pmu K^+ K^- +\pmu \pi^- \pi^++ \pmu K^-
    K^+)+\cdots.
  \end{align}
  This is interesting, since only $\pi$ and $K$ mesons showed up. Using ordinary
  Feynman rules and multiplying in the constants
  \begin{align}
    &A^\mu \pmu \pi^- \pi^+ -A^\mu \pmu \pi^+ \pi^-
    \rightarrow i(\pmp +\qmp) + i\pmp
    \nonumber\\
    &\Longrightarrow \frac{F_0^2}{4} \frac{4ei}{F_0^2}(\qmp+2\pmp)=ei(\qmp + 2\pmp)
    \\
    \nonumber\\
    &A^\mu \pmu K ^- K^+ -A^\mu \pmu K^+ K^-  
    \rightarrow i(\pmk+\qmk) +  i\pmk
    \nonumber\\
    &\Longrightarrow ei(\qmk + 2\pmk).
  \end{align}
\item 
  The same as above, except that $\mu \rightarrow \nu$.
\item
  Here one external field couples to an interior. The second term in
  Eq.~\eqref{eqqcdale1} contain the field couplings needed. The $\pi^0, K^0$ and
  $\eta_8$ is not considered since they did not show up in the three vertex. The
  resulting terms are then
  \begin{align}
    tr(\chi U^\dagger +U\chi^\dagger)= i\sqrt2 F_0 B_0
    (P_{\pi^+}\pi^-+P_{K^+}\pi^-)+\cdots
  \end{align}
  where
  \begin{align}
    i\sqrt2 F_0 B_0P_{\pi^+}\pi^-\rightarrow i\sqrt2 F_0 B_0
    \nonumber\\
    i\sqrt2 F_0 B_0P_{K^+}K^-\rightarrow i\sqrt2 F_0 B_0.
  \end{align}
\item
  The same result as above
\item
  In the propagator the derivative term squared and the mass term are of
  interest. So
  \begin{align}
    &tr(\frac{1}{F_0^2}(\pmu \Phi)^2)+tr(\chi U^\dagger +U\chi^\dagger)=
    \nonumber\\
    &\frac{2}{F_0^2}(2\pmu \pi^+ \partial^\mu \pi^- + 2\pmu K^+ \partial^\mu K^-)
    -\frac{4B_0}{F_0^2}(2\hat{m}\pi^+ \pi^- +\hat{m}K^+ K^-)+\cdots
  \end{align} 
  where $\hat{m}=(m_u+m_d)/2$. The complex fields above propagate in the
  following way
  \begin{align}
    &\pmu \pi^+ \partial^\mu \pi^--2B_0\hat{m}\pi^+\pi^-
    \rightarrow \frac{i}{p_{\pi}^2-2B_0\hat{m}}
    \nonumber\\
    &\pmu K^+ \partial^\mu K^--B_0(\hat{m}+m_s)K^+K^-
    \rightarrow \frac{i}{p_{K}^2-B_0(\hat{m}+m_s)}.
  \end{align}
\item
  Achieve exactly the same terms as above
\item
  Here the resulting terms are
  \begin{align}
    &\frac{i}{(p_{\pi}+q_\pi)^2-2B_0\hat{m}}
    \nonumber\\
    &\frac{i}{(p_K+q_K)^2-B_0(\hat{m}+m_s)}.
  \end{align}
\item
  This is the photon propagator, and only terms including 
  $-\frac{1}{4}F_{\mu\nu}F^{\mu\nu}-\frac{1}{2(1-\xi)}({\partial_\mu}{A^\mu})^2$ are 
  considered. The terms propagate according to
  \begin{align}
     -\frac{1}{4}F_{\mu\nu}F^{\mu\nu}-\frac{1}{2(1-\xi)}({\partial_\mu}{A^\mu})^2 
     \rightarrow -i\int_{0}^{\infty}\frac{d^4q}{(2\pi)^4q^2}(g_{\mu
      \nu}-\frac{\xi q_\mu q_\nu}{q^2}).
  \end{align}
\end{enumerate}

Adding all the parts for the Feynman diagram gives the following amplitudes
for the $\pi$ and $K$ case, respectively:
\begin{align}
  &\int_{0}^{\infty}\frac{d^4q}{(2\pi)^4}\frac{-2F_0^2B_0^2e^2((q_\pi+2p_\pi)^2-\xi 
    \frac{[q\cdot (q_\pi+2p_\pi)]^2}{q^2})}{(p^2_ \pi
    -2B_0\hat{m})^2((p_\pi+q_\pi)^2-2B_0\hat{m})q^2}
  \nonumber\\
  &\int_{0}^{\infty}\frac{d^4q}{(2\pi)^4}\frac{-2F_0^2B_0^2e^2((q_K+2p_K)^2-\xi 
    \frac{[q\cdot (q_K+2p_K)]^2}{q^2})}{(p^2_ K
    -B_0(\hat{m}+m_s))^2((p_K+q_K)^2-B_0(\hat{m}+m_s))q^2}.
\end{align}
In the next step we calculate the one-loop integrals, where we integrate from zero 
to infinity. After transforming from Minkowski space to Euclidian space 
($p^2=(p^0)^2-(\vec{p})^2 \rightarrow -p_E^2=(p^0)^2+(\vec{p})^2$) we can introduce 
a cut-off $\Lambda$ and split the integral into two parts. The first part from zero to 
$\Lambda$ gives us the following integral where the q independent terms are included 
in B
\begin{align}
  B\int_{0}^{\Lambda}\frac{d^4q}{(2\pi)^4}\frac{(q_\pi+2p_\pi)^2-\xi 
    \frac{[q\cdot (q_\pi+2p_\pi)]^2}{q^2}}{((p_\pi+q_\pi)^2-2B_0\hat{m})q^2}.
\end{align}
For the photons with a momentum greater than $\Lambda$ you have to use other methods 
to calculate their contribution (see section 3). The important part is that in the 
end we want to compare the $p^2e^2$ contribution from the
effective QCD Lagrangian in Eq.~\eqref{eqqcdale1} with the $p^2e^2$ part of
Eq.~\eqref{eqntloel1}. This is done by putting e.g $m_q=0$ or $p^2=0$ on both sides, 
i.e. the integrals are greatly simplified.

\section{Generating functionals}

The two following sections are based on Ref. \cite{ab1}.

\begin{figure}[h]
  \begin{center}
    \begin{picture}
      (210,65)(0,25)
      \SetWidth{1.}

      \Line(114,70)(143,58)
      \Line(110,50)(140,50)
      \Line(114,30)(143,42)
      \CArc(155,50)(15,0,360)
      \Line(170,50)(200,50)
      \Line(167,58)(196,70)

      \Line(116,72)(112,68)
      \Line(116,68)(112,72)
      \Line(116,32)(112,28)
      \Line(116,28)(112,32)
      \Line(112,52)(108,48)
      \Line(112,48)(108,52)
      \Line(202,52)(198,48)
      \Line(202,48)(198,52)
      \Line(198,72)(194,68)
      \Line(198,68)(194,72)

      \CArc(196,30)(1,0,360)
      \CArc(194,20)(1,0,360)

      \Text(40,50)[]{$G^{(n)}(x_1,...,x_n)$ =}
      \Text(100,72)[]{$x_1$}
      \Text(98,50)[]{$x_2$}
      \Text(210,70)[]{$x_n$}      
    \end{picture}
  \end{center}
   \caption{The figure shows n external fields coupled to some process.}
  \label{fig:gf1}
\end{figure}
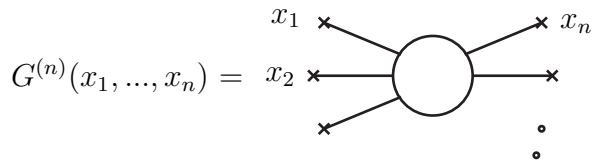

In this thesis Green functions in the form of two- respective four- point
functions are used. They both
originate from the matrix elements describing the physical process and the
general matrix element is connected to a $n$-point function
\begin{align}
  \langle0\mid T(\phi(x_k)\cdots \phi(x_p))\mid 0\rangle
  =i^n\frac{\delta^n\ln W[j]}{\delta j(x_k)\cdots \delta j(x_p)}
  =G^{(n)}(x_1,...,x_n)
\end{align}
where j(x) is the classical source field used to probe the theory, and it can
be thought of as an external field. A physical picture of the above is shown 
in Fig.~\ref{fig:gf1}.
The generating functional W is associated with the sources in the following way
\begin{align}
  \label{eqgl1}
  W[j]=\int{[d\phi]e^{i\int{d^4x(\mathcal{L}(\phi,\partial \phi)-j\phi)}}}.
\end{align}
and one should think of $\mathcal{L}$ in Eq.~\eqref{eqgl1} as describing physical
processes inside the big bubble in Fig.~\ref{fig:gf1} and the external fields as
creating and annihilating particles on the edge of the bubble. So
the bubble contain all relevant Feynman diagrams, e.g the ones in 
Fig.~\ref{fig:ntloel1}.

The two-point function is often written as a Fourier transform into momenta 
of the matrix elements with 
only two sources (four sources for a four point function). In the path
integral formalism the two-point function can be written as
\begin{align}
  \Pi (q^2)=i\int{d^4xe^{iqx}\langle{0\mid T(P(0)P^\dagger (x))\mid 0\rangle}}
\end{align}
where P(x) is a pseudoscalar source in our case.

\section{External fields}

Here the idea about the external field formulation using generating
functionals is presented. The external fields are a method that we use to be able 
to go from the generating functional to the Green function, and they can be thought 
of as the generalized j current of appendix B. One can also use the 
analogy where a particle is accelerated into a target particle and we investigate 
what comes out. The particle in this case is our external field and the target is 
the Green function. 

The use of external fields has some advantages, i.e.
it is independent of the G/H parameterization, allows for a well
defined off-shell amplitude and it is obviously invariant under chiral
transformations. One more advantage is that the connection between QCD and the
local properties become clearer. 

The external fields are added to the Lagrangian and the generating functional 
can be written as
\begin{eqnarray}
  \label{eqef1}
  e^{i\Gamma(v_\mu, a_\mu, s, p)}
  &\equiv&
  \frac{1}{Z}\int{[dq][dG]e^{i\int{d^4x(\mathcal{L}_{QCD}^0+\bar{q}
        \gamma^\mu(v_\mu+a_\mu \gamma_5)q-\bar{q} (s-ip\gamma_5)q)}}}
  \nonumber\\
  &\approx&
  \frac{1}{Z}\int{[dU]e^{i\int{d^4x\mathcal{L}_{eff}(U, v_\mu, a_\mu, s, p)}}}
\end{eqnarray}
where the approximation is valid at low energies as we saw in the case of the 
non-linear
sigma model. As seen in Eq.~\eqref{eqef1}, the external currents contain quark
currents and when the Lagrangians are differentiated with respect to the
external field variables the pure quark current show up. The EM part could
also be added to the Lagrangian above so that EM currents show up. 
\newline

{\Large \bf Acknowledgments}\\
I would like to thank Johan ``Hans'' Bijnens for his valuable advises and for 
answering all my questions. The Master students at Vinterpalatset for their company at 
all the lunches and coffee breaks. And all the people that contributed with rewarding 
discussions, where a special thanks goes out to Timo L\"ahde.

\end{document}